\documentclass[fleqn,10pt]{wlscirep}
\usepackage[utf8]{inputenc}
\usepackage[T1]{fontenc}

\definecolor{myblue}{rgb}{0.0, 0.0, 1.0}

\usepackage{comment}
\usepackage{floatrow}	
\usepackage{graphicx}
\usepackage{subcaption}

\usepackage{times}
\usepackage{csquotes}
\usepackage{graphicx}
\usepackage{framed}
\usepackage{bm}

\global\long\def\Edisplacement{\mathbf{D}}%
\global\long\def\Stress{\boldsymbol{\sigma}}%
\global\long\def\Piezoelectricmodule{\mathbf{B}}%
\global\long\def\Permittivity{\mathbf{A}}%
\global\long\def\elas{\mathbf{C}}%
\global\long\def\momentum{\boldsymbol{\mu}}%

\global\long\def\disp{\mathbf{u}}%
\global\long\def\potential{\phi}%
\global\long\def\ensemble#1{\left\langle #1\right\rangle }%
\global\long\def\rg{\mathbf{W}}%
\global\long\def\willis{\mathbf{S}}%
\global\long\def\tp#1{#1^{\mathsf{T}}}%
\global\long\def\tp#1{#1^{\mathsf{T}}}%
\global\long\def\conj#1{#1^{*}}%
 \def\XXint#1#2#3{{\setbox0=\hbox{$#1{#2#3}{\int}$}  
   \vcenter{\hbox{$#2#3$}}\kern-.5\wd0}}

\def\grad{\nabla} 
\def\sys{\mathbf{r}} 
\global\long\def\lm{\boldsymbol{\mu}}%
\global\long\def\rv{\boldsymbol{G}}%
\global\long\def\tmat{\mathsf{T}}%
\global\long\def\vect#1{\mathbf{#1}}%
\global\long\def\Tmat{\mathsf{\mathsf{T}}}%

\usepackage{hyperref}
\usepackage{bookmark}

\definecolor{erk_pink}{HTML}{DB1C8E}

\hypersetup{
  breaklinks=true,
  colorlinks=true,
  linkcolor=erk_pink,
  filecolor=black,
  urlcolor=erk_pink,
  citecolor=erk_pink,
  pdfborder=0 0 0,
}

\title{Symmetry-Driven Artificial Phononic Media}

\author[1]{Simon Yves}
\author[2]{Michel Fruchart}
\author[3]{Romain Fleury}
\author[4]{Gal Shmuel}
\author[5,6]{Vincenzo Vitelli}
\author[7,*]{Michael R.~Haberman}
\author[1,8,*]{Andrea Al\`{u}}
\affil[1]{Photonics Initiative, Advanced Science Research Center, City University of New York, New York, NY 10031, USA}
\affil[2]{Gulliver, CNRS, ESPCI Paris, Université PSL, 75005 Paris, France}
\affil[3]{Laboratory of Wave Engineering, EPFL, 1015 Lausanne, Switzerland}
\affil[4]{Faculty of Mechanical Engineering, Technion–Israel Institute of Technology, Haifa 32000, Israel}
\affil[5]{University of Chicago, James Franck Institute, 929 E 57th Street, Chicago, IL 60637}
\affil[6]{University of Chicago, Kadanoff Center for Theoretical Physics, 933 E 56th St, Chicago, IL 60637}
\affil[7]{Walker Department of Mechanical Engineering, The University of Texas at Austin, Austin, Texas 78712, USA}
\affil[8]{Physics Program, Graduate Center, City University of New York, New York, NY 10026, USA}
\affil[*]{Corresponding authors: aalu@gc.cuny.edu; haberman@utexas.edu}

\begin{abstract}
Phonons are quasiparticles associated with mechanical vibrations in materials. They are at the root of the propagation of sound and elastic waves, as well as of thermal phenomena, which are pervasive in our everyday life and in many technologies. The fundamental understanding and control of phonon responses in natural and artificial media are key in the context of communications, isolation, energy harvesting and control, sensing and imaging. It has recently been realized that controlling different symmetry classes at the microscopic and mesoscopic scales in synthetic media offers a powerful tool to precisely tailor phononic responses for advanced acoustic and elastodynamic wave control. In this Review, we survey the recent progress in the design and synthesis of artificial phononic media, namely phononic crystals and metamaterials, guided by symmetry principles. Starting from tailored broken spatial symmetries, we discuss their interplay with time symmetries for non-reciprocal and non-conservative phenomena. We also address broader concepts that combine multiple symmetry classes to induce exotic phononic wave transport. We conclude with an outlook on future research directions based on symmetry engineering for the advanced control of phononic waves.
\end{abstract}
\begin{document}

\flushbottom
\maketitle

\thispagestyle{empty}

\section{Introduction}

\begin{figure}[t]
\begin{minipage}[t]{0.5\columnwidth}%
\begin{center}
\includegraphics[scale=.85]{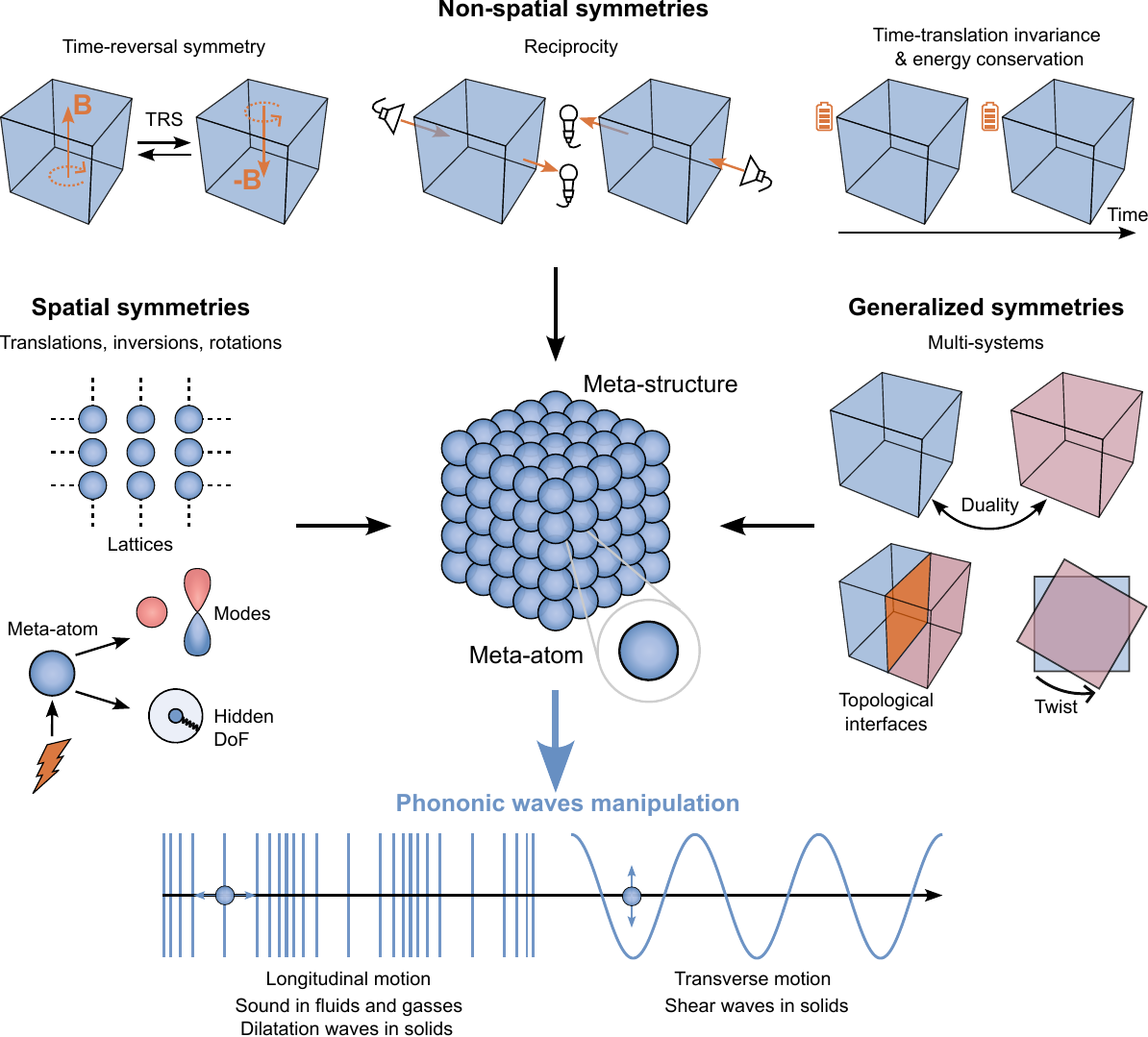}
\par\end{center}
\end{minipage}\hfill{}%
\begin{centering}
\centering{}
\par\end{centering}
\caption{\textbf{Symmetry-driven artificial phononic media.} Identification and selective breaking of the various symmetries characterizing phononic materials and meta-structures, both at the microscopic and macroscopic scales. These broken symmetries enable enhanced control over acoustic and elastodynamic wave propagation. In this review we focus on different symmetry classes, namely spatial symmetries and non-spatial symmetries the latter of which includes reciprocity, time-reversal and time-translation invariance symmetry, energy conservation, and the generalization of these symmetries to families of systems.}
\label{fig:FigIntroSymm}
\end{figure}

As outlined by Pierre Curie in the late 19th century\cite{curie1894symetrie}, symmetries play a fundamental role in the physical understanding of various natural phenomena, especially in the context of wave behavior. As a consequence, physicists and engineers
have been hunting for the presence or absence of symmetries in systems to not only understand, but also tailor their physical
properties. In this work, we provide a unified review and perspective on the latest research progress studying the propagation of mechanical waves in meso-structured materials, such as phononic crystals and metamaterials, under the paradigm of symmetries and symmetry breaking. 
Recent reviews of acoustic and elastic metamaterials and metasurfaces~\cite{deymier2013acoustic,ma2016acoustic,cummer2016controlling,Assouar2018natrev,craster2023mechanical} feature exciting advances in the control of mechanical wave propagation through structured media. Here, we aim at articulating this progress across different classes of symmetry-engineering, showing how this approach can provide a powerful perspective to understand, design and optimize phononic meta-structures, as well as to offer insights for the development of the next generation of phononic devices.

Overall, the concept of symmetry is very general: it includes any transformation that keeps an object unchanged.
This is the case for the familiar spatial symmetries, which we discuss after introducing some basic phononics concepts. We then expand on to abstract symmetries connected to the temporal evolution of a system including reciprocity, time-reversal, time-translation invariance, and energy conservation.
Finally, we discuss generalized symmetries involving families of systems such as dualities and twist symmetries (Fig.~\ref{fig:FigIntroSymm}).

\section{Symmetry-driven phononics in a nutshell}

\subsection{Mechanical waves and how they propagate}

This work focuses on elastic waves (mechanical waves in solids) and acoustic waves (mechanical waves in fluids and gases). Possibly the simplest description of mechanical waves is from the perspective of continuum theories~\cite{Truesdell1960,Truesdell2004,Landau6,Landau7}. 
For instance, acoustic waves in a simple fluid can be captured by the acoustic wave equation
\begin{equation}
    \beta \partial_t^2 p = \rho_0^{-1} \nabla^2 p,
    \label{acoustic_intro}
\end{equation}
where $p(t, \vect{r})$ is the pressure, $\rho_0$ is the density of the unperturbed fluid and $\beta$ its isentropic compressibility (the inverse of the isentropic bulk modulus $B = \beta^{-1}$).
This equation can be obtained by linearizing the Navier-Stokes equations along with an isentropic equation of state~\cite{Pierce,Landau6}.

Similarly, continuum elasticity describes the propagation of elastic waves in an ideal, uniform, and isotropic solid by the equation of motion~\cite{achenbach1973wave,Landau7}
\begin{equation}
    \rho \partial_t^2 \vect{u} 
    = 
    \mu \nabla^2 \vect{u}
    +
    (B+\frac{\mu}{3}) \nabla (\nabla \cdot\vect{u}), 
    \label{elastic_intro}
\end{equation}
in which $\rho$ is the mass density of the solid, $\disp(t, \vect{r})$ is the displacement field that measures the motion of the solid with respect to a reference configuration, while $\mu$ and $B$ are the shear and bulk moduli, respectively.
Both Eqs.~\eqref{acoustic_intro} and \eqref{elastic_intro} can be traced to conservation laws that arise from symmetry: the conservation of mass $\partial_t \rho + \rho \nabla\cdot\vect{v} = 0$ and the conservation of linear momentum $\partial_t\lm = \nabla \cdot \Stress + \bf{f}$
in which $\lm = \rho \vect{v}$ is the density of linear momentum, $\vect{v}=\partial_t\disp$ is the velocity field, $\Stress$ is the stress tensor, and $\bf{f}$ is the density of body forces applied externally (set to zero here).
Acoustic waves are longitudinal, while elastic waves include both longitudinal compression waves and transverse shear waves.
Both acoustic and elastic waves have a polarization (the direction of the oscillating velocity or displacement field, respectively), which can be, in the case of plane waves, related to their longitudinal or transverse nature\cite{Bliokh2025}.

Equations~\eqref{acoustic_intro} and \eqref{elastic_intro} may not be sufficient to describe wave propagation. This can happen when (\textit{i}) the material does not satisfy the required symmetries, (\textit{ii}) there are other degrees of freedom in the system, (\textit{iii}) the system exhibits nonlinearities, or (\textit{iv}) a continuum theory is not appropriate given the size of material heterogeneities of the medium when compared to the wavelength. Several of these conditions may apply at the same time, especially in the context of engineered materials.

\subsection{Materials with lower symmetries}
\label{less_symmetries}

When mechanical waves propagate in materials with lower symmetries, the propagation equations typically acquire additional terms accounting for new couplings between degrees of freedom that arise when the symmetry of the system is reduced. It can happen that the equations keep roughly the same form, but are more complicated. 
For instance, a more general version of Eq.~\eqref{elastic_intro} that does not assume isotropy reads (neglecting body forces)
\begin{equation}
    \rho \partial_t^2 u_i 
    = 
    \partial_j[ C_{i j k \ell}  \partial_\ell u_k ]
    \label{elasticity_tensor}
\end{equation}
in which $u_i$ is the $i$th Cartesian component of $\vect{u}$ and $\partial_i \equiv \partial/\partial r_i$ are partial derivative with respect to space.
Here, $C_{i j k \ell}$ is a material property known as the elasticity tensor that relates the stress $\sigma_{ij}$ to the displacement gradients $\partial_\ell u_k$ through $\sigma_{ij} = C_{i j k \ell} \partial_\ell u_k$, in the same way as Hooke's law for a linear spring.
Compared to Eq.~\eqref{elastic_intro}, more terms are present, but they are roughly of the same shape (a second-order derivative with respect to space). Any remaining symmetry is encoded in $C_{ijk\ell}$ \footnote{In particular, spatial symmetries represented by matrices $U \in O(d)$ constrain the elastic tensor through $C_{i j k \ell} = R_{i i'} R_{j j'} R_{k k'} R_{\ell \ell'} C_{i' j' k' \ell'}$. But other kinds of symmetries can be broken, see for instance Sec.~\ref{odd_elasticity}.}.

It can also happen that new kinds of terms appear. As an example, consider sound waves propagating in a moving fluid flowing along the $x$ axis.
In the lab frame, these can be described by the equation
\begin{equation}
    \beta \partial_t^2 p = \rho_0^{-1} \nabla^2 p - 2 \beta v_0 \partial_x \partial_t p - \beta v_0^2 \partial_x^2 p
    \label{movingfluid}
\end{equation}
in which $v_0$ is the velocity of the unperturbed fluid in the lab frame. 
The term with mixed time and space derivatives $\partial_x \partial_t p$ in Eq.~\eqref{movingfluid} is known as a \emph{Willis coupling term}. 
As we shall see in Secs.~\ref{breaking_inversion} and \ref{breaking_reciprocity}, it is the consequence of having broken the inversion symmetry of the system and time-reversal invariance.
The $\partial_x^2 p$ term is the consequence of the anisotropy of the system; contrary to the Willis coupling, it could arise in a mirror-symmetric system.

\subsection{Additional degrees of freedom in the medium}
\label{other_dofs}

The continuum description of Eqs.~\eqref{elastic_intro} and \eqref{acoustic_intro} focuses on the displacement field $\vect{u}(t,\vect{r})$. 
However, this may not be enough to encode all the relevant degrees of freedom in the system. 
As an example, mechanical degrees of freedom can be coupled with heat transport (in thermoelasticity) or with electromagnetism (in piezoelectricity and electrostriction~\cite{Gerhard2016}).
Typically, the key feature of piezoelectric crystals is the conversion of mechanical energy into electricity and back. As such, it is necessary to include the electromagnetic field in the description of phononic materials when piezoelectric phenomena emerge (see Sec.~\ref{piezoelectricity_electromomentum_coupling}).
This example illustrates the interplay between symmetries and relevant degrees of freedom: inversion symmetry of bound charge distribution within the medium must be broken for piezoelectricity to occur in order to induce strain-dependent electric dipoles. When this is not the case, the mechanical and electrical degrees of freedom are decoupled and can be treated interdependently.
Similarly, one must consider magnetoelastic effects in naturally occurring magnetic materials\cite{Lee1955} and magnetoelastic metamaterials\cite{Lapine2011} via magnetic order parameters\cite{Reid2018,DuTremoletDeLacheisserie2005}.
In the same vein, the polarization of acoustic waves requires consideration of the acoustic velocity field, not only the scalar pressure field (Sec.~\ref{angular_momentum}).

Besides, additional mechanical degrees of freedom not captured by the instantaneous displacement field can play a role.
In the simplest situations, these features can be captured by viscoelastic\cite{lakes2009viscoelastic} or elastoplastic\cite{liu2024harnessing,wallen2025strongly} models in which the deformation history affects the evolution of the deformation in a rate-dependent way, so the history of the displacement gradient $\nabla\vect{u}(t,\vect{r})$ must be included in the constitutive models.
Alternatively, some aspects of microstructural deformation asymmetries may be retained in the constitutive description as is the case of micromorphic elasticity \cite{maugin2010mechanics}, of which micropolar (Cosserat) elasticity theories which include microscale rotations and couple stresses are a special case \cite{Cosserat1909,Muhlhaus1995,Toupin1962,Mindlin1964,maugin2010mechanics,Ehlers2020}.
These can be necessary to describe metamaterials where complex unit cells lead to internal motions (known as non-affine deformations) that may deviate considerably from the (macroscopic) average.
For instance, consider a half-filled bottle of water. 
If the bottles are not transparent, we cannot track the motion of the water, so the relationship between the total linear momentum and any directly observable displacement becomes non-trivial\cite{milton2007modifications}.

When the system is linear, several simplifications arise. In particular, we can perform a Fourier transform in space and time, and this allows us to hide degrees of freedom at the price of having complex-valued \enquote{nonlocal} material coefficients that depend on the wavevector $\vect{q}$ and the frequency $\omega$ and therefore represent a convolution in space and time. The convolution in time, in particular, represents a form of time-translation invariant memory in the system \footnote{To see how this works in a simplified case, consider two degrees of freedom $a$ and $b$ and say that $\dot{a} = \alpha a + \beta b$ while $\dot{b} = \gamma a + \delta b$. Then, in Fourier space, $i \omega a = \alpha_{\text{eff}}(\omega) a$ in which $\alpha_{\text{eff}}(\omega) = \alpha + \beta \gamma/[i\omega - \delta]$.}. For instance, Eq.~\eqref{elasticity_tensor} becomes
\begin{equation}
    \rho \omega^2 u_i(\bm{q}, \omega)
    = 
    C_{i j k \ell}(\bm{q}, \omega) q_j q_\ell u_k(\bm{q}, \omega).
    \label{elasticity_tensor_fourier}
\end{equation}
The case of Eq.~\eqref{elasticity_tensor} would correspond to a constant $C_{i j k \ell}$ not depending on the wavevector $\vect{q}$ and the frequency $\omega$.
In fact, the material coefficients $C_{i j k \ell}(\omega)$ of conventional materials such as steel or air display a weak dependency on frequency due to microscale behavior such as molecular relaxation processes and internal friction \cite{Lakes1998,lakes2009viscoelastic}. 
From the perspective of rheology, this means that all materials have a viscous response in addition to an elastic response.
In principle, this Fourier transform procedure allows us to eliminate many degrees of freedom, but keep in mind that we still need to keep track of the degrees of freedom we care about because we can experimentally manipulate or measure them, like the electric field in piezoelectricity.
Equation~\eqref{elasticity_tensor_fourier} is deceptively simple: if we Fourier transform back to real time, the product of Fourier transforms becomes a convolution of the instantaneous strain and the relaxation function of the material\cite{lakes2009viscoelastic}, which makes the presence of memory explicit in the system.
When the system is nonlinear, these simplifications do not hold and a case-by-case study is required. 
However, symmetries still constrain the possible nonlinear terms in the equations and can be harnessed to control the behavior of the system~\cite{bertoldi2017flexible,Lapine2014}.

\subsection{Beyond continuum theories}
\label{continuum_or_else}

In some situations, it may be necessary to describe the material as a collection of discrete units coupled to each other (for instance, a collection of masses connected by springs).
This is the case, for instance, when an artificial material is physically constructed out of weakly-coupled individual units such as resonators, whose size is not negligible when compared with the typical scale of spatial variations of the phononic wave propagating in the system.
In some situations, a continuum description may also be insufficient even if we include many degrees of freedom~\cite{Kevrekidis2002,Lapine2014}.

In the context of wave propagation, the most common way of describing an assembly of coupled resonators or modes is temporal coupled mode theory\cite{Fan2003,WonjooSuh2004} whereby the evolution of the complex amplitude $a_n(t)$ of the resonant mode $n$ is described by (see also BOX2)
\begin{equation}
    \partial_t a_m = \sum_{n} L_{m n} a_n
    \label{cmt}
\end{equation}
The matrix (or operator) $\hat{L}$ with components $L_{m n}$ represents the coupling between different modes (the operator $\hat{H} = i \hat{L}$ is often called a Hamiltonian by formal analogy with quantum mechanics).
In this case, the symmetry of the system arises from the interplay between the symmetry of the modes and of the geometry of their couplings, and it is encoded in operators acting in the same space as $\hat{L}$.
As shown in Fig.~\ref{fig:FigIntroSymm}, and analogously to atomic orbitals, modes can be scalar, vectorial (in which case $a_n$ are the components of the modes), etc. They can be further organized into a metamaterial with a crystalline structure, allowing one to design its symmetries on demand.
In this coupled mode theory, transformations of the degrees of freedom (that may or may not be symmetries) are encoded in invertible operators $\hat{U}$ acting in the same space as $\hat{L}$.
The transformation $\hat{U}$ is a symmetry when it commutes with $\hat{L}$, i.e.~when $\hat{U} \hat{L} = \hat{L} \hat{U}$.

\subsection{Getting a continuum theory from a mesoscopic description}
The techniques to derive a continuum theory from a collective description of the individual elements of the material are known as coarse-graining, averaging, or homogeneization~\cite{Pavliotis2008}. 
In metamaterials described by linear equations of motion, homogenization can be achieved at the price of having frequency- and momentum-dependent coefficients (see Sec.~\ref{other_dofs}), which introduces nonlocalities into the homogenized description of the material.
In the general case, however, approximations are required to remove irrelevant degrees of freedom by exploiting a separation of timescales using methods such as adiabatic elimination of averaging~\cite{Kuramoto1984,Kunihiro2022}. 
For instance, spatially periodic media, which are invariant under discrete spatial translations, can be treated using Bloch theory, while random media, which are on average translation-invariant, can be described using disorder-averaging techniques.
Correspondingly, the relation of the continuum fields to the microscopic degrees of freedom (which can be appropriately chosen combinations or disorder averages and so on) may change even though their physical meaning should always be the same.
We refer to \cite{Pavliotis2008,Willis2011PRSA,born1956dynamical,Nassar2015,Nassar2016,Craster2024,Admal2010,Srivastava2015} for more details including techniques and methods: 
Refs.~\cite{born1956dynamical,Lutsko1989,Irving1950} deal with interacting point particles; Refs.~\cite{scheibner2020odd,Fruchart2020,Poncet2022} for extend to cases with less symmetries; 
Refs.~\cite{Norris2012,Nassar2015,Nassar2016,Meng2018,Willis2009} describe Bloch-Floquet techniques for spatially periodic media, while \cite{willis1981variational,Milton07,WILLIS2012MRC} describe ensemble average methods for disordered media, a review can be found in Ref.~\cite{Srivastava2015}.

\subsection{Elastic solids: a case study}
\label{elastic_case_study}

We now consider a generalization of Eq.~\eqref{elastic_intro} that describes the propagation of elastic waves in solids known as Willis materials~\cite{Norris2012}, and that will serve as a case study throughout the review. Other examples and generalizations will be described as they arise.

The equation of motion of elastic waves in a solid takes the form 
\begin{subequations}
\label{willis_case_study}
\begin{equation}
    \partial_t \mu_i = \partial_j \sigma_{ij} + f_i.
\end{equation}
in which $\mu_i$ is the density of linear momentum, $\sigma_{ij}$ is the stress tensor, and $f_i$ is the density of external body forces.
In addition, we consider the constitutive relations
\begin{equation}
\begin{pmatrix}
\Stress\\
\momentum
\end{pmatrix}=\begin{pmatrix}
\elas & \willis \\
\tilde{\willis} & \boldsymbol{\rho}
\end{pmatrix}\begin{pmatrix}
\nabla{\disp}\\
\partial_t\disp
\end{pmatrix}
\quad
\text{or in index notation}
\quad
\begin{split}
    \sigma_{ij} &= C_{ijk\ell} \partial_\ell u_{k} + S_{ijk} \partial_t u_k
    \\
    \mu_i &= \tilde{S}_{ik\ell} \partial_\ell u_{k} + \rho_{ij}  \partial_t u_j
\end{split}
\label{eq:EffConstRel}
\end{equation}
\end{subequations}
according to which the momentum density $\mu_i$ and the stress $\sigma_{ij}$ are proportional to both the time and space derivatives of \emph{some} displacement field $\disp(t, \bm{r})$, that may or may not be the displacement of the center of mass. This is in stark contrast with the behavior of conventional solids discussed in the introduction for which these constitutive relations are uncoupled.

In the constitutive relations~\eqref{eq:EffConstRel}, the mass density $\bm{\rho}$ is no longer a scalar field, but a rank-two tensor that arises because the displacement field may not coincide with the displacement of the center of mass, $\elas$ is a rank-four elastic tensor, and the third-order tensors $\willis$ and $\tilde{\willis}$ known as Willis couplings\cite{willis1981variational,WILLIS2012MRC,Milton07} can be seen as the phononic analogue of bianisotropic tensors in electromagnetism\cite{milton06cloaking,koo2016acoustic,Sieck2017prb,FIETZ2010pysicaaB} whose local version is rooted in spatial asymmetry (\ref{Willis couplings}).
As discussed in the introduction, the quantities in Eq.~\eqref{eq:EffConstRel} may be effective quantities that have to be properly defined on a case-by-case basis.
In addition, all quantities in \eqref{eq:EffConstRel} may depend on the frequency $\omega$ and the wavevector $\bm{q}$ (e.g. $\sigma_{ij}$ is $\sigma_{ij}(\omega, \bm{q})$), making them non-local in space and time: the products in \eqref{eq:EffConstRel} represent a convolution in space and time.
In this case, the material coefficients like $C_{ijk\ell}$ may be complex-valued to encode phase lags between stress and strain at a material point. 
Typically, Willis coupling coefficients are connected to a weak form of nonlocality \cite{willis1981variational,WILLIS2012MRC}, due to the fact that the dynamic effective response of the medium depends on both the local response of the material point and its interaction with neighboring heterogeneities, described by the gradients of the phononic fields.
In the following, we define \enquote{local} materials as those that are adequately described by linear response coefficients that do not depend on $\bm{q}$ (i.e. in the $\bm{q} \to \bm{0}$ limit). This generally occur if the microstructure of the medium is sufficiently small relative to the wavelength, leading the constitutive response at any material point to depend only on the fields at that point.


\section{Breaking spatial symmetries}
\label{breaking_spatial_symmetries}

A bulk material, viewed from the continuum perspective, is invariant under all spatial translations, rotations, and inversions. {These are collectively known as isometries and form the Euclidean group $\mathbb{E}^d$ (where $d$ is the dimension of space).} 
These symmetries underpin constraints and conservation laws.
For instance, translation and rotation symmetries lead to linear and angular momentum conservation, respectively~\cite{KosmannSchwarzbach2011}, while point group symmetries guarantee that quantities with different symmetries are decoupled~\cite{BradleyCracknell,Malgrange2014,dresselhaus2007group}. 
This makes their selective breaking an efficient tool to engineer wave propagation within artificial media, which is the focus of this section.

\subsection{Breaking translation symmetry}
\label{Breaking translation symmetry}

Inhomogeneities in a medium, such as a spatial interface between two materials, break spatial translation symmetries. 
In such a system, the conservation of the physical momentum $\momentum$ that underlies the wave equations still holds, because it is related to the joint translation of the medium and the wave through Noether's theorem. 
In contrast, the translation of the waves alone is not a symmetry because the medium is inhomogeneous. 
As a consequence, another quantity called the wave momentum is no longer conserved\cite{Maugin2015,Maugin1993,Stone2002}.
Intuitively, this can be seen from the fact that at an interface, the refracted and reflected waves carry different wavevectors compared to the incident field. The spatial repetitions of such interfaces can result into band foldings (Sec.~\ref{phononic_crystals}) or more complex structures (Sec.~\ref{beyond_phononic_crystals}).
Scattering at single interfaces can also be designed to manipulate waves in both near and far field (Sec.~\ref{Interfaces}). 
These aspects are the focus of this section.

\subsubsection{Phononic crystals}
\label{phononic_crystals}

In a phononic crystal~\cite{deymier2013acoustic}, continuous translation invariance is broken, but the system remains invariant under a set of discrete translations collected in a group called a Bravais lattice~\cite{Ziman1979}.
This breaking of continuous translation invariance also partially break rotation and reflection symmetries: the remaining symmetries are captured in mathematical objects called space groups \cite{BradleyCracknell,dresselhaus2007group,Malgrange2014}.
For instance, one could consider a version of Eq.~\eqref{elasticity_tensor} in which the elasticity tensor $\elas(\bm{r})$ and the density $\rho(\bm{r})$ depend on the position $\bm{r}$ in a spatially periodic fashion, e.g. $\rho(\bm{r}) = \rho(\bm{r} + a \hat{\vect{e}})$ in which $a \hat{\vect{e}}$ is a vector defining the discrete periodicity of the phononic crystal.

In order to analyze such a system, we use Bloch-Floquet theory~\cite{Brillouin1946,Ziman1979,Simon2013,BradleyCracknell}.
In a nutshell, the spatial periodicity $\vect{r} \to \vect{r}+a\hat{\vect{e}}$ implies that the plane waves $e^{i \bm{q}\cdot\bm{r}}$ and $e^{i(\bm{q}+\rv)\cdot\bm{r}}$ where $\rv\cdot a\hat{\vect{e}}=2\pi n, n\in\mathbb{Z}$, are indistinguishable, and so one can define the wavevector $\bm{q}$ on a reduced region called a Brillouin zone that has periodic boundary conditions. 
The wavevector space is therefore tiled with copies of a \enquote{first} Brillouin zone centered around $\vect{q}=0$.
The resulting wave propagation is captured by a band structure, consisting of a set of dispersion relations $\omega_i(\bm{q})$ (and the corresponding vibrational modes) that are repeated periodically in the wavevector space outside the first Brillouin zone\footnote{Despite the mathematical equivalence between the different Brillouin zones in an infinite periodic phononic crystal, the internal structure of vibrational modes (without the plane wave envelope) can be better understood by assessing in which Brillouin zones it is supported, i.e.~how fast it spatially oscillates in various directions. This gives information on the behavior of the system when an interface or a defect is present.}.
This description encompasses and goes beyond the metamaterial picture in which an effective continuum theory with modified material constants is used to describe the behavior of the system probed at long wavelengths\cite{Laude2009,deymier2013acoustic,Srivastava2015}. It can perturbatively be seen as the result of \emph{folding} the dispersion relation of waves in a homogeneous medium into the first Brillouin zone, which can be further harnessed to control wave propagation by considering families of symmetries, see Sec.~\ref{nonlocality_engineering}.

One of the key features of phononic crystals is that they can exhibit band gaps, i.e.~frequency bands where no wave propagation can occur.
Mathematically, this can be understood using the so-called transfer matrix $\mathsf{T}(\omega,E,\rho)$ that describes the propagation of waves with frequency $\omega$ along a given direction through a finite region by relating the wave amplitudes on the left to that on the right\cite{Laude2009,Dwivedi2016,Schomerus2017,Markos2008}.
For instance, the transfer matrix corresponding to the one-dimensional version $\rho \partial_t^2 u = \partial_x \sigma$ with $\sigma = E \partial_x u$ is given by\cite{Thomson1950,Banerjee2018}
\begin{equation}
    \begin{pmatrix}
        u_R \\ 
        \sigma_R
    \end{pmatrix}
    = 
    \Tmat
    \begin{pmatrix}
        u_L \\
        \sigma_L
    \end{pmatrix}
    \qquad
    \text{with}
    \qquad
    \Tmat = \begin{pmatrix}
        \cos \vartheta & Z^{-1} \sin \vartheta \\
        - Z \sin \vartheta & \cos \vartheta
    \end{pmatrix}
    \label{transfer_matrix_example}
\end{equation}
in which $L/R$ mean left/right and where $Z = \rho c^2 \kappa$ is the impedance of the medium of dispersion relation $\omega = c \kappa$ with $c^2 = E/\rho$, and $\vartheta = \kappa h$.
In a lossless medium, the eigenvalues of the transfer matrix are, in general, of the form $e^{\pm i q_B}$ and the dispersion relation of the bands is given by $\text{tr}[\Tmat(\omega)] = 2 \cos q_B$ in which $q_B$ is the (dimensionless) Bloch wavenumber, which is real-valued for propagating (Bloch) states and purely imaginary for non-propagating states\footnote{When $u$ and $\sigma$ have more than a single component, $\tmat$ is a larger matrix and these expressions have to be adjusted accordingly.}.
For the particular $\Tmat$ in Eq.~\eqref{transfer_matrix_example}, we find that $q_B(\omega) = \vartheta = \omega/c h$, so we recover the bulk dispersion relation.
Consider now a phononic crystal obtained by alternating two media with different impedances $Z_{1,2} \equiv Z (1 \pm \epsilon)$, corresponding to a step-wise variation of $E(\bm{r})$  and/or $\rho(\bm{r})$.
The transfer matrix corresponding to the unit cell is $\tmat_{\text{uc}} = \tmat_1 \tmat_2$ [$\tmat_i$ is obtained from $\tmat$ in Eq.~\eqref{transfer_matrix_example} by replacing $Z$ with $Z_i$]. 
As $\text{tr}[\tmat_{\text{uc}}(\omega)]/2 = [\epsilon^2 - \cos(2 \vartheta)]/(\epsilon^2 - 1)$, the inequality $|\text{tr}\,\tmat | \leq 2$ does not always hold. 
When it is not the case, the Bloch wavenumber $q_B$ is not real-valued: the corresponding frequencies $\omega$ correspond to the band gap in which waves cannot propagate.
The critical frequencies $\omega^*$ at which solutions change from propagating (in the bands) to non-propagating (in the gap) are called band edges, and correspond to exceptional points of the transfer matrix\cite{Novikov1984,Figotin2005,Saha2023,Saha2025}.

As a consequence, exceptional points can arise in the complex wavevector space even in passive lossless media.
It is also possible to design them by exploiting the co-existence of multiple wave polarizations in planar elastodynamics.
For instance, in elastic laminates with isotropic constituents, the polarization conversion between shear and dilatation waves can induce exceptional points even away from the edges of the Brillouin zone\cite{Lustig2019,MOKHTARI2020ijes}. These conservative laminates can give rise to anomalous wave phenomena such as negative refraction\cite{Lustig2019} and beam steering\cite{Lustig2019,MOKHTARI2020ijes}. By including an anisotropic component, it is also possible to break the symmetry of leftward and rightward waves and excite axially frozen modes with a finite transmittance despite a vanishing axial group velocity\cite{FISHMAN2024JMPS}. Going back to dispersion relations, it turns out that all the band structures of 1D layered systems are encapsulated in a compact universal manifold (Fig.~\ref{fig:Spatial}a for the two-layer case discussed above), which depends only on the impedance mismatches (not on the volume fraction of the constituents nor their specific physical properties) and from which it is possible to calculate the density of the gaps in the spectrum\cite{Shmuel2016JMPS,LUSTIG2018jmps}. In Fig.~\ref{fig:Spatial}a, the green part of the torus corresponds to gaps, its boundary to the band edges, and the band structure is constructed by wrapping a line (in blue) around the torus.

The band folding approach can also be employed at the sub-wavelength scale in the presence of locally resonant elements with a strong albedo close to resonance, as demonstrated by the structure-induced negative refraction of sound in crystalline metamaterials made of soda cans~\cite{kaina2015negative,lemoult2016soda} (Fig.\, \ref{fig:Spatial}b). In parallel, the existence and size of band gaps are also constrained by the symmetry of the system~\cite{Yablonovitch1993,Watanabe2018}.
Overall, the domain of phononic crystals led the development of various advanced mechanical properties including bandgaps for strong field confinement\cite{maldovan2013sound}, waveguiding\cite{khelif2004guiding} and focusing\cite{yang2004focusing}. 
Band gaps have also been observed in internal gravity waves, a class of mechanical wave present in stratified fluids like the Oceans\cite{Ghaemsaidi2021}. Besides, phononic crystals play a major role in the control of elastic polarizations, ranging from bandgaps with elastic polarization selectivity\cite{ma2016polarization} to multi-physics interactions in cavity optomechanics where strong interactions between photons and phonons can be obtained in phoxonic structures acting as dual photonic-phononic crystals \cite{djafari2016phoxonic}.

\subsubsection{Broken translation invariance beyond phononic crystals}
\label{beyond_phononic_crystals}
Phononic crystals (with their discrete translation invariance) are not the only way to produce structures that have broken translation symmetry while remaining uniform bulk materials in some sense. 
For instance, quasi-periodic, amorphous, hyperuniform, or disordered systems can be seen as generalizations of phononic crystals.
In these systems, slightly different approaches are required. 
Disordered and aperiodic systems cannot be handled using Bloch theory as there is no translation invariance at all.
However, these systems still have some regularity which can be handled using noncommutative geometry, in which Fourier space are replaced by a mathematical object called a $C^*$-algebra\cite{connes1995noncommutative,Bellissard1992,Bellissard2014}.
This has been applied to phononic topological states\cite{Prodan2009,Apigo2019,Mitchell2018} that are discussed in Sec.~\ref{Interfaces}.
{Quasi-crystalline phononic structures, which can be seen as projections of higher-dimensional periodic structures in which the usual symmetry-based approach can be used, have also been extensively studied\cite{beli2021mechanics} for their bandgap properties\cite{lai2002large}, waveguiding\cite{marti2021edge,davies2022symmetry} broadband asymmetric transmission\cite{li2014broadband}, topological pumping\cite{ni2019observation,rosa2021exploring} and fractal rainbow trapping\cite{davies2023graded}, as shown in Fig.\, \ref{fig:Spatial}c.}
In all of these systems, non-local couplings can become important and affect wave propagation (see Sec.~\ref{nonlocality_engineering}).

\subsubsection{Engineering interfaces}
\label{Interfaces}
The boundary of a medium is the most extreme case of spatial translation symmetry breaking. It is also an essential part in defining wave-matter interactions: the boundary permits to interact with the bulk. 
In order to engineer phononic devices, we can either try to perform impedance matching in order to minimize the role of interfaces as much as possible, or embrace them as an engineering knob. 
As an example, metasurfaces are two-dimensional metamaterials which allow one to both control surface waves in near field and to perform beam shaping in far field by controlling the structure and symmetries of the surface\cite{Assouar2018natrev}. 

Another approach consists in harnessing boundary states that can exist at the interface between a medium and air or between two media. The existence of such interface states can be captured from a scattering perspective\cite{Laude2009,Dwivedi2016,Schomerus2017,Markos2008} (BOX2), a description related to the transfer matrix approach mentioned in Sec.~\ref{phononic_crystals}. When two good mirrors are placed face to face, they form a resonant cavity, where standing waves can be maintained until they are damped by losses of some kind, and whose resonant interaction with the environment is described by a scattering matrix $S(\omega)$ (Fig.\, \ref{fig:Spatial}d, left panel). These resonant modes are obtained by requiring that a round-trip in the cavity leaves a wave in phase with itself. In other words, the dephasing $\Delta\phi$ picked during the round-trip should be a multiple of $2\pi$. For a cavity of size $L_c$, $\Delta\phi=2k(\omega)L_c+\phi_L+\phi_R$ where $k(\omega)$ is the dispersion relation of the medium in the cavity, and $\phi_{L/R}$ the reflection phases on the left/right sides. Now, let us consider a cavity where the walls are replaced with a phononic crystal. For frequencies $\omega$ in a band gap, the phononic crystal acts as a frequency-dependent mirror with reflection phases $\phi_{L/R}(\omega)$ arising from the multiple interferences on the Bragg planes of the crystal. The solutions $\omega^*$ of $\Delta\phi(\omega)=0 \; [\text{mod }2\pi]$ in the limit where $L_c=0$ correspond to edge states (also known as Tamm states, see \cite{kaliteevski2007tamm,xiao2014surface,levy2017topological}). These edge states arise at the interface between the left and right phononic crystals, thereby acting as a virtual cavity (Fig.\, \ref{fig:Spatial}d, right panel). This can be extended to cases where one of the media could be vacuum or a boundary condition. It turns out that in some selected instances, the presence of interface states can be traced to the existence of nontrivial topological invariants in the bulk\cite{hasan2010colloquium}. 
We refer the reader to previous reviews for details\cite{ma2019topological,huber2016topological,miniaci2021design,zhang2018topological,xue2022topological,ni2023topological,yves2022topologicalRev,shankar2022topological,coulais2021topology,ding2022non,zhu2023topological} and to Sec.~\ref{generalized_symmetries} for the relation between topology and symmetries beyond translations.
Let us emphasize that this "bulk-boundary correspondence" is not always valid\cite{souslov2019topological,tauber2019bulk,tauber2020anomalous,gangaraj2020physical}, while at the same time alternative approaches to define bulk topology suggest that the origin of certain non-symmetry-protected edge states may still be traced to the bulk\cite{Zhong2024a,Zhong2024b}.

\subsection{Breaking inversion symmetry}
\label{breaking_inversion}
Systems with inversion symmetry preserve the spatial symmetry or anti-symmetry of wavefield profiles, such as monopoles or dipoles, respectively. These can be related to different phononic physical quantities, making inversion symmetry breaking a good design strategy for generalized bi- and  tri-anisotropic phononic media (BOX1). This is the focus of this subsection.

\subsubsection{Willis coupling}\label{Willis couplings}

In Sec.~\ref{elastic_case_study}, we have introduced Willis coupling, which is described by the rank-three tensors in Eq.~\eqref{eq:EffConstRel}.
If inversion symmetry ($\bm{r} \to -\bm{r}$) is present, the Willis coupling tensor $S_{ijk}(\omega, \bm{q})$ in Eq.~\eqref{eq:EffConstRel} must satisfy $S_{ijk}(\omega, \bm{q}) = - S_{ijk}(\omega, -\bm{q})$ (the same is true for $\tilde{S}$).
Hence, there is no Willis coupling in an inversion symmetric local material (i.e. for $S_{ijk}(\omega, \bm{q} \to 0)$).
{This can be understood from the fact that the $S_{ijk}$ relates a vector and a second order tensor, which do not have the same symmetries}.
Conversely, purposely breaking the inversion symmetry of the elastic impedance\cite{Sieck2017prb,rps20201wm,liu2019willis} within metamaterials is a good design strategy to induce enhanced Willis couplings. As an experimental example, Fig.\,\ref{fig:Spatial}e (top panel) shows the unit cell of an elastic structured beam made of resonant meta-atoms whose inversion symmetry is broken\cite{liu2019willis}. This directly results in a Willis coupling that relates the momentum ($\mu_z$) and strain ($\partial_x{u_z}$) within the medium.
In addition to the elastic case, breaking inversion symmetry also yields Willis couplings in the context of longitudinal sound propagating in fluids\cite{Sieck2017prb,Melnikov2019nc,quan2018prl,merkel2018unidirectional,peng2022fundamentals,esfahlani2021homogenization,Willis2009,Lau2019}. For instance, such acoustic Willis couplings have been evidenced experimentally by using a subwavelength asymmetric scatterer in a one-dimensional impedance tube measurement\cite{Muhlestein2017nc} (Fig.\,\ref{fig:Spatial}e, bottom panel). 

We can gain insight on the microscopic origin of Willis coupling by sending a sound wave on a subwavelength scatterer (for which $ka \ll 1$ in which $a$ is the size of the scatterer and $k$ is the wavenumber). 
When the object is mirror-symmetric, it scatters a monopole field $M_{A}$ as a response to the local pressure $p$ and a dipole field $\boldsymbol{D_{A}}$ as a response to the local velocity $\boldsymbol{v}$ field, which is captured by a polarizability tensor (inset of Fig.\,\ref{fig:Spatial}f).
When the scatterer breaks mirror-symmetry with respect to the direction of incidence, however, both the pressure and the velocity contribute to both the monopolar \textit{and} dipolar scattered fields. 
These cross-polarizabilities correspond to the scattering version of Willis couplings\cite{quan2018prl}, and can lead to strong differences in the backward scattering from waves impinging from opposite directions\cite{sounas2014extinction}. 
The forward scattering remains the same as long as reciprocity holds (Section \ref{breaking_reciprocity}).
Breaking inversion symmetry within resonant scatterers can lead to cross-polarizabilities of the same order as the diagonal ones\cite{quan2018prl}, like in the split ring (Helmholtz resonator) in Fig.\,\ref{fig:Spatial}f.
These can serve as building blocks for metameterials exhibiting strong macroscopic Willis couplings\cite{Sieck2017prb},
whose asymmetric acoustic responses in reflection are relevant to wavefront shaping for sound\cite{Lawrence2020JASA,li2019highly,li2018nc,esfahlani2021homogenization} and elastic waves\cite{liu2019willis,hao2022experimental,chen2023controlling}, as well as for particle manipulation\cite{sepehrirahnama2022willis}.

\subsubsection{Piezoelectricity and electromomentum coupling}
\label{piezoelectricity_electromomentum_coupling}

Combined with the presence of electric charges, spatial inversion symmetry breaking allows one to couple mechanical and electric fields. This electromechanical coupling is a feature of piezoelectric materials, which is directly related to the broken centro-symmetry of their atomic structure\cite{Mason1950fk}. 
It is evidenced by the change in electric polarization in the material upon mechanical strain. 
In turn, the inverse piezoelectric effect corresponds to the generation of stress in response to an electric field. 
Accordingly, the stress in a piezoelectric material is related to the gradient of the displacement field and the gradient of the electric potential field via the relation $\Stress=\elas:\grad\disp+\tp{\Piezoelectricmodule}\cdot\grad\potential$, where $\Piezoelectricmodule$ is the piezoelectric coupling tensor in stress-charge form and $\potential$ is the electric potential. Similarly, the electric displacement field $ \Edisplacement$ in a piezoelectric medium is also a function of the same fields through the relation $\Edisplacement=\Piezoelectricmodule\cdot\grad\disp-\Permittivity\cdot\grad\potential$ where $\Permittivity$ is the dielectric permittivity tensor.  

Breaking inversion symmetry from the viewpoint of the piezoelectric properties of the medium leads to an emergent constitutive coupling between electrostatics and dynamics, as was first shown using source-driven continuum homogenization\cite{PernasSalomon2019JMPS}, and thereafter using retrieval methods\cite{rps20201wm} and discrete models\cite{Muhafra2023PRApplied}.
In these systems, the electric polarization and velocity fields are coupled, as well as the linear momentum and the electric field, by the so-called electromomentum couplings\cite{PernasSalomon2019JMPS} (Fig.\,\ref{fig:Spatial}g). As a result, each one of the kinetic fields (stress, linear momentum density, electric displacement) depends on all three kinematic fields (strain, velocity, electric field) (BOX1), and therefore such materials are called tri-anisotropic materials.
In symbolic matrix notation, the constitutive relations of electromomentum-coupled materials take the form
\begin{equation}
\begin{pmatrix}
{\Stress}\\
{\momentum}\\
{\Edisplacement}\\
\end{pmatrix}=\begin{pmatrix}
\elas & {\willis} &  \tilde{{\Piezoelectricmodule}} \\
{\tilde{\willis}} & {\boldsymbol{\rho}} & {\tilde{\rg}} \\
{\Piezoelectricmodule}  & {\rg} & -{\Permittivity}\\
\end{pmatrix}\begin{pmatrix}
{\grad\disp}\\
{\partial_t{\disp}}\\
{\grad\potential}\\
\end{pmatrix},\label{eq:EffConstRelEM}
\end{equation}
where $\tilde{\rg}$ and $\rg$ are the second-order electromomentum coupling tensors\cite{muhafra2021}. 
Like the Willis couplings \cite{Sieck2017prb} and the magnetoelectric couplings\cite{alu2011restoring}, 
the electromomentum couplings are required to ensure that the constitutive relations satisfy physical constraints \cite{pernassalomn2020prapplied}.
Unlike Willis couplings, the electromomentum effect depends on the circuit conditions, yielding an intrinsic electrical tunability for wave manipulation\cite{rps20201wm,Danawe2023APL,HUYNH2023EML}, and scattering\cite{lee2022maximum}. 
In addition, recent works analyzing the polarizability of electromomentum-coupled scatterers\cite{lee2022maximum,wallen2022polarizability} have shown that the polarizability tensor must consider both electric and magnetic field scattering, due to the time-varying nature of all fields, and that the polarizabilities coupling mechanical and electric fields can reach the same magnitude as the diagonal terms, even when the Willis coefficients vanish\cite{rps20201wm}.

\subsection{Breaking rotation symmetry}
\label{breaking_rotation}
Breaking rotation symmetry permits to engineer artificial media beyond isotropic constraints, thereby providing a tensorial richness to their mechanical response (BOX1). Rotational symmetry of the underlying medium also underpins the conservation of the angular momentum of the waves, which makes its breaking an efficient tool to control the chirality of phononic fields. These aspects are the focus of this subsection.

\subsubsection{Anisotropy engineering}
\label{breaking_isotropy}
Isotropic materials are endowed with full rotation symmetry. In these systems, like Eqs.~\ref{acoustic_intro} and \ref{elastic_intro}, waves propagate in all directions in the same way. 
In contrast, anisotropic materials do not have full rotation invariance. 
When they are still homogeneous, the remaining symmetries are contained in a mathematical object called a point group\cite{BradleyCracknell,Malgrange2014}.
As a consequence of anisotropy, material properties must be encoded in (anisotropic) tensors, such as the elastic tensor $\elas$ in Eq.~\eqref{elasticity_tensor}. This has important implications on wave propagation\cite{cummer2016controlling,ma2016acoustic,Christensen2012Aniso}. As an example, consider a 2D collection of subwavelength resonators consisting of circular cavities carved in a rigid medium, which individually host a mass connected to the boundary by springs\cite{milton2007modifications,oudich2014negative} (Fig.\, \ref{fig:Spatial}h, left panel).
In such a system, the measurable displacement field does not necessarily coincide with the displacement of the center of mass, because the internal masses may be hidden.
As a consequence, the velocity $\bm{v}$ associated with the displacement field and the linear momentum density $\bm{\mu}$ are related through $\bm{\mu} = \boldsymbol{\rho} \bm{v}$ through an effective mass density tensor of the form
\begin{equation}
\bm{\rho}
=
\begin{pmatrix}
\rho_{xx} & \rho_{xy}\\
\rho_{yx} & \rho_{yy}\\
\end{pmatrix}
\label{eq:Effrhotensor}
\end{equation}
whose components can either be positive or negative, corresponding to an in-phase or out-of-phase macroscopic response of the medium, respectively\cite{milton2007modifications}.
Acoustic waves in such a 2D anisotropic system can be described by a generalization of Eq.~\eqref{acoustic_intro} in which\cite{ma2016acoustic} $\beta \partial_t^2 p = \partial_i ([\bm{\rho^{-1}}]_{ij} \partial_j p) = 0$
%
in which $\rho^{-1}$ is the matrix inverse of $\rho$.
In a coordinate system where the mass density tensor is diagonal, the resulting dispersion relation is
\begin{equation}
    \omega^2 = 
    \frac{q_x^2}{\beta\,\rho_{xx}}
    +
    \frac{q_y^2}{\beta\,\rho_{yy}}
\end{equation}
This expression directly shows that the anisotropy affects the shape of the isofrequency contours of the system, which describe the spatial properties of wave propagation in the medium at the operating frequency $\omega$ in Fourier space. 
Indeed, depending on the relative signs of the eigenvalues of the mass density tensor, the isofrequency contours may have different topologies (open or closed), with fundamentally different consequences on wave behavior\cite{krishnamoorthy2012topological}.
As a landmark example of open contour topology, phononic hyperbolic metamaterials support extremely anisotropic properties, such as broadband, diffraction-free directional ray-like propagation, negative refraction, and enhanced wave-matter interaction{\cite{poddubny2013hyperbolic,gomez2016flatland,huo2019hyperbolic}}. 
These features have been evidenced experimentally for sound using membranes in a 2D waveguide\cite{shen2015broadband} (Fig.\,\ref{fig:Spatial}h, right panel). Other acoustic implementations have been demonstrated in the context of hyperlenses \cite{li2009hyperlens,lu2012hyperlenses}, as well as in elastodynamics using patterned plates\cite{oh2014truly,zhu2016single,lee2016extreme,dong2018broadband} and asymmetric pillars\cite{yves2022moire}.
Following recent advances in nano-optics\cite{gomez2016flatland}, sonic hyperbolic metasurfaces have also been proposed\cite{quan2019hyperbolic}. 
Rotation symmetry can be broken further in the case of multi-layer phononic structures, leading to extremely anisotropic and reconfigurable wave propagation, that we discuss in the context of twistronics in Sec.~\ref{twistronics}.

\subsubsection{Controlling the angular momentum of waves}
\label{angular_momentum}

\begin{figure}[t]
\vspace*{-1.5cm}
\begin{minipage}[t]{0.5\columnwidth}%
\begin{center}
\includegraphics[scale=.85]{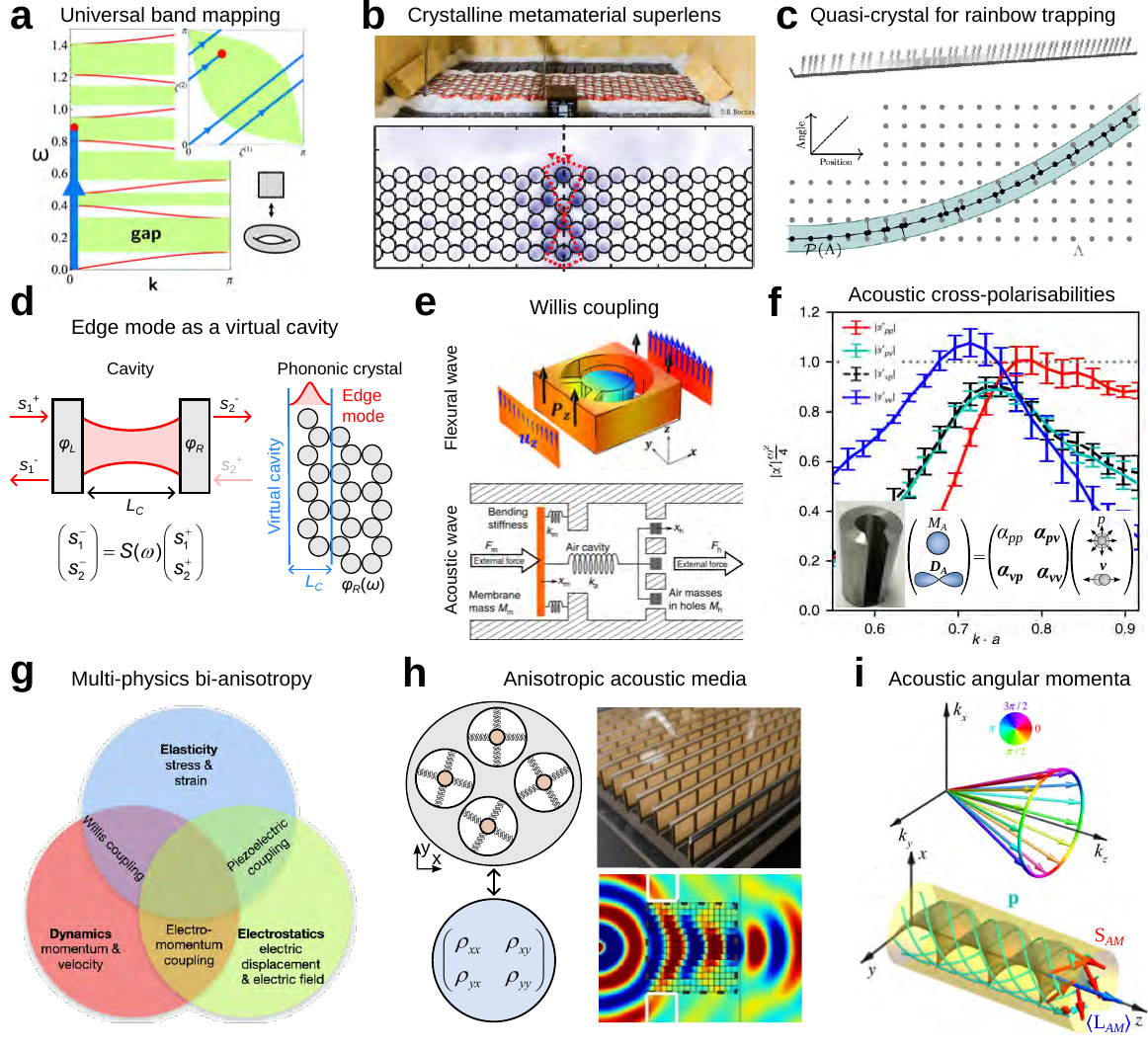}
\par\end{center}
\end{minipage}\hfill{}%
\begin{centering}
\centering{}
\par\end{centering}
\caption{\textbf{Phononic phenomena induced by breaking spatial symmetries.} \textbf{a}, Universal torus (square with opposite edges identified in inset) onto which all infinite band diagrams of 1D phononic crystals are mapped (example in the main panel). The frequency $\omega$ acts as a time-like parameter defining a linear flow on the torus $\overrightarrow{\zeta}(\omega)=\omega\left(h_1/c_1,h_2/c_2\right)\,\text{mod}\ \pi=:\left(\zeta^{(1)},\zeta^{(2)}\right)$, such that different crystals are mapped to flows of  different slopes $\zeta^{(2)}/\zeta^{(1)}$; the gap region (green) is universal for crystals with identical impedance mismatch. \textbf{b}, Subwavelength-scaled crystal made of hollow soda cans for negative refraction and superlensing of acoustic surface waves (the dashed red lines show the corresponding ray tracing). \textbf{c}, Quasi-crystalline phononic lattice ($\Lambda$) stemming from an effective projection onto a quadratic curve ($P(\Lambda)$) that is graded along its length, leading to fractal rainbow trapping. \textbf{d}, Correspondence between a conventional resonant cavity made of two mirrors with reflection phases $\mathrm{\phi_{L,R}}$, described by a scattering matrix $S(\omega)$ (left) and the virtual cavity at the edge of a gapped phononic crystal, whose reflection phase $\mathrm{\phi_R(\omega)}$ depends on the operating frequency, hosting a boundary mode. \textbf{e}, Willis coupling between momentum $P_z$ and strain $\partial_x{u_z}$ for flexural waves in a structured beam (top), and  schematics of an asymmetric scatterer responsible for air-borne acoustic Willis coupling  (bottom). \textbf{f}, Experimental verification of maximal off-diagonal polarizabilities within an asymmetric Helmoltz resonator (inset)  which match the conventional monopole and dipole polarizability. These maximal off-diagonal polarizabilites are scattering versions of acoustic Willis coupling. \textbf{g}, Diagram showing the multi-physics couplings that exist in a trianisotropic medium: Willis coupling, piezoelectricity and electro-momentum coupling. The local part of these interactions stems from different symmetry-breakings  at the subwavelength scale (see BOX1). \textbf{h}, (Left) general model of an acoustic medium whose macroscopic effective response yields a dynamic mass density tensor. (Right) acoustic metamaterial with membranes along one spatial direction, leading to hyperbolic wavefront propagation. \textbf{i}, Acoustic Bessel beam with a rotating phase profile (top) responsible for helicoidal wavefronts, known as a vortex beam, carrying non-zero integral orbital angular momentum $\mathrm{\ensemble{\boldsymbol{L_{AM}}}}$ and spin angular momentum density $\mathrm{\boldsymbol{S_{AM}}}$ (bottom). 
Panel (\textbf{b}) adapted with permission from \cite{lemoult2016soda}. Springer Nature Limited.
Panel (\textbf{c}) adapted with permission from \cite{davies2023graded}. American Physical Society. 
Panel (\textbf{e}) (top) reproduced with permission from\cite{liu2019willis}. American Physical Society. (Bottom) reproduced with permission from\cite{Muhlestein2017nc}. Springer Nature Limited.
Panel (\textbf{f}) adapted with permission from\cite{Melnikov2019nc}. Springer Nature Limited.
Panel (\textbf{g}) reproduced with permission from\cite{pernassalomn2020prapplied}. American Physical Society.
Panel (\textbf{h}) (right) adapted with permission from\cite{shen2015broadband}. American Physical Society.
Panel (\textbf{i}) adapted \phantom{with with} with permission from \cite{bliokh2019spin}. American Physical Society.
}
\label{fig:Spatial}
\end{figure}

In the same way a phononic wave field carries energy and linear momentum, it can also carry angular momentum $\bm{J} = \boldsymbol{L_\mathrm{AM}} + \boldsymbol{S_\mathrm{AM}}$ which can be decomposed into an orbital angular momentum $\boldsymbol{L_\mathrm{AM}}$ and a spin angular momentum $\boldsymbol{S_\mathrm{AM}}$\cite{Bliokh2025}. 
The orbital angular momentum $\boldsymbol{L_\mathrm{AM}} = \sys\times\momentum$ is associated with rotations of spatial patterns in the wave field, and typically manifests as helicoidal wave fronts, as shown in Fig.\,\ref{fig:Spatial}i (top panel) in the case of an acoustic Bessel beam\cite{bliokh2019spin}. 
The spin angular momentum is associated with rotations of the polarization associated to the vector part of the wave field. 
In acoustic waves, this means that one has to take into account the velocity field in addition to the scalar pressure field\cite{Jones1973,shi2019observation,burns2020acoustic,Bliokh2025}. 
The corresponding spin density $\boldsymbol{S_\mathrm{AM}} = \text{Im}(\rho _0\boldsymbol{\conj{v}}\times\boldsymbol{v})/2 \omega$, where $\rho_0$ is the density of the fluid and $\bm{v}$ its velocity field, vanishes upon integration over the entire medium in homogeneous media, but it can be nonzero in the presence of field inhomogeneities \cite{bliokh2019spin,burns2020acoustic} and for surface waves\cite{bliokh2019transverse,long2020symmetry,long2023universal}. 
Hence, both elastic\cite{long2018intrinsic,bliokh2022elastic} and acoustic waves can carry a spin angular momentum (Fig.\,\ref{fig:Spatial}i, bottom panel).

Based on this, one can directly manipulate the angular momentum of phononic waves by engineering the breaking of rotation invariance of the propagating medium. For instance, it is possible to generate $\boldsymbol{L_\mathrm{AM}}$ both in acoustics and elastodynamics by using spiral-shaped tube sections\cite{chaplain2022elastic,chaplain2022elastic2} or with multiple sources with tailored phase delays\cite{wang2018topological}. One can also implement such rotation symmetry breaking within the resonant structure of metasurfaces\cite{jiang2016convert,fu2020sound,gao2021emitting,fan2021acoustic,jiang2020modulation,zhang2023topologically} to induce helicoidal wavefronts. Besides, the reduced planar rotation symmetry within some topological phononic crystals can be related to modes exhibiting a local orbital angular momentum which can induce vortices in the far field\cite{wang2021vortex,lu2016valley} (see Section \ref{symmetry_to_topology}).
Breaking rotation invariance can also lead to spin-momentum locking, in which the direction of linear momentum determines (locks) the direction of $\boldsymbol{S_\mathrm{AM}}$\cite{shi2019observation}, leading to spin-dependent propagation and selective wave routing at waveguide intersections\cite{long2020realization}. 

Besides, a phononic spin-orbit coupling between orbital and spin angular momenta can be induced by a mismatch between the rotation symmetries of the unit cell and the lattice. 
For instance, the elastic version of this behavior has been obtained in  micro-structured mechanical materials\cite{frenzel2017three,frenzel2019ultrasound} which twist in a specific direction when pushed along their axis, and proposed in quasi-crystals\cite{chen2020isotropic}. 
In acoustics, spin-orbit coupling emerges in metastructures of dipolar modes with twisted inter-cell couplings, resulting in chirality-induced negative refraction\cite{wang2021spin}.
These ideas have been applied in the context of imaging\cite{noetinger2023superresolved}, multiplexing\cite{shi2017high} and particle manipulation\cite{baresch2016Observation,melde2016holograms,cox2019acoustic,melde2023compact}.

\section{Breaking non-spatial symmetries}
\label{breaking_nonspatial_symmetries}

Usual phononic materials and their idealized versions exhibit several non-spatial symmetries which are tightly connected to the specificities of the temporal dimension: reciprocity, time-reversal invariance, time-translation invariance, and energy conservation. 
These symmetries, which are related to each other (BOX2) impose strong constraints on the behavior of waves. Therefore, the combined breaking of spatial and non-spatial symmetries provides numerous opportunities for advanced phononic wave engineering, which is the focus of this section. 

\subsection{Breaking reciprocity}
\label{breaking_reciprocity}

Spatial asymmetries can tailor the reflection and absorption of waves from opposite sides of a material, but are not sufficient to produce a genuine asymmetry in the transmission of waves between two points in space, like one would have in a diode~\cite{Maznev2013}. 
As a wave propagates through the material, it undergoes a phase shift proportional to the distance traveled and an amplitude change due to the presence of losses. 
Usually, these modifications do not depend on whether the wave travels to the right or to the left, even in a inhomogeneous and arbitrarily shaped medium\cite{sounas2014extinction}.
The reason for that is a discrete non-spatial symmetry known as reciprocity, that relates incoming and outgoing waves in a scattering process (BOX2).
Non-reciprocal phononic media, where this symmetry is broken, can asymmetrically transmit mechanical energy, with potential applications to information and heat transport.

In practice, most ways of breaking reciprocity entail breaking spatial symmetries as well as time-reversal invariance ($t \to - t$), another fundamental discrete non-spatial symmetry which corresponds to reversing the flow of time, like watching the dynamics backward as in a movie played with a reversed sequence of images.
From the perspective of wave propagation, time-reversal invariance is related to the microscopic reversibility of the medium where waves propagate, which is violated by the presence of external biases such as magnetic fields or rotation.
Yet in systems where energy is not conserved (Sec.~\ref{breaking_energy_conservation}), it is in principle possible to have non-reciprocal systems that are time-reversal-symmetric\cite{Buddhiraju2020,Guo2022}.
We refer the reader to Refs.\cite{nassar2020nonreciprocity,zangeneh2019active,caloz2018electromagnetic,Asadchy2020,Guo2022,Maznev2013} for precise definitions and detailed discussions on reciprocity in wave propagation.

\subsubsection{Breaking reciprocity through an external bias}
\label{breaking_reciprocity_external_bias}

A common way of breaking reciprocity is to impose an external bias whose sign reverses under time-reversal symmetry, such as a magnetic field. Microscopic reversibility implies that symmetry of the transmission coefficients is
preserved only by reversing the bias when flipping the propagation direction. 
On the contrary holding the bias constant when interchanging source and receiver positions typically leads to a nonreciprocal response\cite{cummer2014selecting}.
This works well in electromagnetism\cite{caloz2018electromagnetic} because the propagation of light in certain materials is strongly influenced by available magnetic fields. 
Doing so is also possible with phonons, for instance through magnetoacoustic couplings\cite{Luthi2005,kittel1958interaction}, but the effect is often weak.
For instance, nonreciprocity induced by magnetoacoustic effects has been predicted and observed in surface acoustic waves\cite{Lewis1972,Heil1982}, and can be strongly  enhanced by the coupling with nonreciprocal spin waves\cite{Verba2018,Verba2019,Shah2020}. 
Alternatively, one can do so through other external biases such fluid flow~\cite{Godin1997,Brekhovskikh1999,Morse1968,Silbiger2022}, rotation\cite{roux1997aharonov}, or odd/Hall viscosity (Sec.~\ref{non_reciprocal_continuum}).

In this context, the use of resonant components enables strong nonreciprocal effects with flow speeds much lower than the speed of sound. 
For example, Fleury \textit{et al.} created a highly non-reciprocal acoustic circulator for audible sound using a ring cavity with a fluid spinning at a fraction of the speed of sound\cite{fleury2014sound} (Fig.\, \ref{fig:Reciprocity}a).
In the absence of an external bias, the two lowest-frequency modes in the circular cavity of Fig.~\ref{fig:Reciprocity}a are degenerate with frequency $\omega_0$. 
These correspond to clockwise and counterclockwise propagating waves, for which the pressure field takes the form $p_\pm(r, \phi) \sim e^{\pm i \phi}$ in polar coordinates, that can be combined into standing waves.
When a background velocity field $v$ is imposed (e.g. with fans), the frequencies of these modes undergo a Doppler shift $\Delta \omega_\pm/\omega_0 \sim \pm v/c$, leading to an acoustic Zeeman effect\cite{fleury2014sound,fleury2015breaking} that lifts the degeneracy. 
The resulting system can be captured by the coupled modes equation 
$\dot{a}_{\pm} = \left(i \omega_{\pm} - \gamma_\pm \right) a_\pm + W_{\pm, \alpha} s^{\text{in}}_\alpha$
in which $W_{\pm, \alpha} = \sqrt{2 \gamma_\pm/3} \exp(\mp 2 \pi/3 (\alpha-1))$ represents the coupling to the three equispaced channels in Fig.\, \ref{fig:Reciprocity}a.
Applying Eq.~\eqref{Seff} gives the corresponding scattering matrix. In the simplified case where $\gamma_\pm = \gamma$ and the system is excited at $\omega = \omega_0$, we find that the transmissions between the channels 1 and 2 are
\begin{equation}
  T_{1\rightleftarrows2} = |S_{12/21}|^2 = \frac{4 \gamma^2 \left(\gamma \pm \sqrt{3} \Delta\omega \right)^2}{9 \left(\gamma^2+(\Delta\omega)^2\right)^2}
\end{equation}
from which we see that (i) $T_{1\to2} \neq T_{2\to1}$ when $\Delta \omega \neq 0$, meaning that the system is not reciprocal, (ii) it is possible to tune the $\Delta \omega$ to have $T_{1\to2} = 0$.

This design was later used as a basis for theoretical and experimental investigations in topological acoustics in which the rings are put on a lattice\cite{khanikaev2015topologically,ding2019experimental,Zhang2021}, for non-reciprocal wave manipulation in the context of Janus metasurfaces\cite{zhu2021janus} and to create non-reciprocal mode conversion in an elastic waveguide\cite{goldsberry2022nonreciprocity}.
Alternatively, it is possible to have the fluid flow by itself if it is made of self-propelled active components \cite{Jorge2024,Yang2020,Souslov2017,shankar2022topological}. 

It is also possible to create non-reciprocal Willis couplings (Sec.~\ref{non_reciprocal_continuum}) in spatially symmetric systems through external biases whose sign reverses under time-reversal symmetry.
Going back to the example of Sec.~\ref{breaking_inversion}, the normalized cross-polarizabilities $\boldsymbol{\alpha}_{vp}$ and $\boldsymbol{\alpha}_{vp}$ relating the acoustic monopole $M_{A} = \alpha_{pp} p + \boldsymbol{\alpha}_{pv} \boldsymbol{v}$ and dipole $\boldsymbol{D_{A}} = \alpha_{vv} \boldsymbol{v} + \boldsymbol{\alpha}_{vp} p$ to the pressure and velocities are also constrained by reciprocity, which imposes\cite{quan2018prl} $\boldsymbol{\alpha}_{vp} = - \boldsymbol{\alpha}_{pv}^T$. 
This constraint can be lifted by a constant bias\cite{quan2021odd}, as in the spatially symmetric acoustic resonator embedded with a rotating flow in Fig.\,\ref{fig:Reciprocity}b.  
In a lossless system, this leads to ${\boldsymbol{\alpha}}_{vp}={\boldsymbol{\alpha}}_{pv}^T$ (a situation refereed to a scattering version of odd Willis coupling in Ref.~\cite{quan2021odd}), resulting in a different power extinguished by the scatterer when excited from the left or from the right, in stark contrast with the reciprocal cross-polarizabilities obtained through inversion-symmetry-breaking alone (Sec.~\ref{breaking_inversion}).

\subsubsection{Breaking reciprocity by combining spatial asymmetries and nonlinearity}
\label{Spatial asymmetry and nonlinearity}

In nonlinear systems, it is possible to break reciprocity dynamically, without an external bias\cite{liang2010acoustic,blanchard2018non}, for instance by combining spatial asymmetries with a medium whose properties depends on the amplitude of the wave. 
Putting a lossy material on one side of the nonlinear medium effectively makes the wave interact with a different medium when excited from either side, making the transmission direction-dependent, i.e., nonreciprocal. 
This simple mechanism has been implemented and studied fundamentally\cite{sounas2017time,darabi2019broadband,devaux2019acoustic,liang2010acoustic,liang2009acoustic}, together with other nonlinear schemes based on related phenomena, such as phononic band gaps\cite{Liu2015,Cui2018}, frequency conversion\cite{boechler2011bifurcation,meng2024roton,guo2023observation}, self-demodulation\cite{devaux2015asymmetric,devaux2019acoustic}, prestretched linkages\cite{Fang2021} or hysteresis\cite{librandi2021programming}. 
The combination of spatial asymmetry and nonlinearity is a common ingredient in all these schemes, as showcased by the non-reciprocal phononic wave transmission stemming from the intrinsic nonlinear acoustic radiation pressure happening at an interface between water and air\cite{devaux2019acoustic} (Fig. \,\ref{fig:Reciprocity}c).

\subsubsection{Non-reciprocal continuum phononic media}
\label{non_reciprocal_continuum}

In the case of continuum media, reciprocity can be framed as a link between the two different excitations and the corresponding responses. 
This constraint is known as Maxwell-Betti or Lorentz reciprocity (BOX1 of Ref.~\citen{nassar2020nonreciprocity}).
In our case study~\eqref{willis_case_study}, it implies that\cite{Muhlestein20160Prsa2}
\begin{equation}
    \label{reciprocity_elastic}
    C_{ijk\ell} = C_{k\ell ij}
    \qquad
    \tilde{S}_{ijk} = S_{jki}
    \qquad
    \rho_{ij} = \rho_{ji}.
\end{equation}
When the elastic tensor is real-valued, breaking its major symmetry ($C_{ijk\ell}\neq C_{k\ell ij}$) also requires to violate energy conservation; this is discussed in Sec.~\ref{odd_elasticity}.
The same holds for the mass density tensor: active metamaterials with a non-reciprocal $\rho_{ij}$ have been realized using feedback loops\cite{wu2023active}.
This is not the case of Willis coupling, for which the constraint due to energy conservation is different\cite{Muhlestein20160Prsa2}, see also\cite{peng2022fundamentals} for acoustic waves.
As mentioned in Sec.~\ref{less_symmetries}, one of the simplest way, mathematically, to obtain Willis coupling is to consider sound waves in a moving fluid, for which we obtain an energy-conserving non-reciprocal bi-anisotropic coupling\cite{quan2019nonreciprocal}. 
Starting from the linearized conservation of mass ($\beta \partial_t p = - \partial_x v$) and of linear momentum ($\rho_0 \partial_t v = - \partial_x p$) that combine into Eq.~\eqref{acoustic_intro}, and performing a Galilean boost ($\partial_t \to \partial_t - v_0 \partial_x$) to account for the motion of the fluid at speed $v_0$, we end up with
\begin{equation}
    \begin{pmatrix}
        \partial_x v \\
        \partial_x p
    \end{pmatrix}
    =
    -
    \begin{pmatrix}
        \beta & \xi \\
        \zeta & \rho_{0}
    \end{pmatrix}
    \begin{pmatrix}
        \partial_t p \\
        \partial_t x
    \end{pmatrix}
    + \mathcal{O}(v_0^2)
\end{equation}
in which $\xi = \zeta = v_0 \beta \rho_0 + \mathcal{O}(v_0^2)$ is cast as purely non-reciprocal acoustic Willis couplings.
A similar feature occurs in colloidal solids driven by a flow\cite{Beatus2006,Beatus2012,Poncet2022}. Nonreciprocal Willis couplings have also been discussed in passive systems such as moving fluids in zero-index metamaterials\cite{quan2019nonreciprocal}.
To induce strong nonreciprocal Willis effects going beyond the limitations imposed by passivity\cite{cho2021acoustic,wen2023acoustic}, active mechanisms such as electronic feedback loops\cite{zhai2019active, chen2020active}, spatio-temporal modulation\cite{nassar2017modulated}, or thermoacoustic amplifiers\cite{olivier2021nonreciprocal} have also been considered.

\medskip

In (meta)fluids and viscoelastic media, non-reciprocal responses can also be encoded in viscosities that violate reciprocity known as odd/Hall viscosities\cite{avron1998odd,fruchart2023odd} corresponding to phonon Hall viscosities\cite{heidari2019hall,barkeshli2012dissipationless,Flebus2023} in the context of phonons in solids.
The Navier-Stokes equations take the general form
\begin{equation}
    \rho (\partial_t + \bm{v} \cdot \bm{\nabla}) v_i 
    = 
    - \partial_i p + \partial_j \sigma_{ij}
    \quad
    \text{where}
    \quad
    \sigma_{ij} = \sigma_{ij}^{\text{h}} + \eta_{ijk\ell} \partial_\ell v_k
\end{equation}
is the stress tensor, split into a hydrostatic and a viscous part. 
The viscosity is encoded in a tensor $\eta_{ijk\ell}$, and reciprocity imposes that $\eta_{ijk\ell}=\eta_{k\ell ij}$.
This constraint is for instance broken in chiral fluids made of actively-spinning components and in magnetized polyatomic gases and plasma, see Ref.\cite{fruchart2023odd} for a review.
Consequences include non-reciprocal wave propagation\cite{Avron1998,han2021fluctuating,soni2019odd,souslov2019topological,sone2019anomalous,Khain2022,Markovich2024,Hosaka2021,Hosaka2023}, which can affect turbulence in the nonlinear regime\cite{deWit2024} as well as topologically protected boundary modes\cite{Souslov2017,sone2019anomalous,shankar2022topological} and non-linear shock waves in compressible fluids\cite{han2021fluctuating,soni2019odd,souslov2019topological} (Fig.\,\ref{fig:Reciprocity}d) and  biological tissues\cite{Chen2025} (see Fig.~\ref{fig:Reciprocity}e).
These are typically manifested in transverse responses, for instance in the velocity field along the piston in a shock (color in Fig.~\ref{fig:Reciprocity}d) or edge flows (red arrows in Fig.~\ref{fig:Reciprocity}e) that are a response to the flows towards the middle (blue arrows) due to the proliferation of cells in tissues.


\subsection{Breaking time-translation invariance}
\label{breaking_time_translation_invariance}
Breaking time-translation invariance entails changing the physical properties of a system in time, so that wave excitations effectively see a different system or medium at different times. 
These time-dependent modulations require an active drive, which may inject or remove energy from the system. 
They can be designed to tune spatial and non-spatial symmetries of the system almost at will, at the price of a higher complexity and cost in design and operation compared to static systems.  
This approach permits to adapt ideas related to the breaking of spatial translation invariance (Sec.\,\ref{Breaking translation symmetry}), although the specific nature of the temporal dimension and its ties to causality provide additional physical constraints that results in fundamentally different wave phenomena. In the case of periodic time-modulations, time-translation invariance remains in a discrete form, yielding the emergence of frequency harmonics. Single temporal interfaces results in different laws of refraction and reflection stemming from the conservation of momentum rather than frequency as in their spatial counterparts. These aspects of phononic time-varying media are reviewed in this section.

\subsubsection{Time-periodic modulations}

While the time dependence can be arbitrary, theoretical and practical investigations have largely focused on periodic temporal modulations, because these are easier to handle. 
These driven systems are called Floquet systems\cite{floquet1883equations,rahav2003effective,goldman2014periodically,rudner2020floquet}, and sometimes \enquote{time crystals}; we emphasize that this is a distinct concept as time crystals in statistical physics~\cite{Khemani2019,Zaletel2023,Yao2020,Avni2025} and biology~\cite{Izhikevich2007,Winfree2001}.

In Floquet systems, the dynamics are described by a periodically driven Hamiltonian $H(t)$ obeying $H(t)=H(t+nT_0)$ for any integer $n$, where $T_0$ is the modulation period.
In contrast with time-independent systems in which the linear response only occurs at the excitation frequency $\omega$, periodically driven systems may exhibit a response at harmonic components $\omega + n \omega_0$ ($n \in \mathbb{Z}$) spaced by $\omega_0\equiv 2\pi/T_0$ \cite{sambe1973steady}, which are referred to as sidebands.
Mathematically, the Floquet theorem\cite{floquet1883equations} decomposes the evolution operator $U(t) = V(t) e^{i t H_{\text{eff}}}$ associated with $H(t)$ into a $T_0$-periodic micromotion $V(t) = V(t+T_0)$ and a long-time evolution described by a time-independent effective Hamiltonian $H_{\text{eff}}$, called the Floquet Hamiltonian.
For instance, Fig.\, \ref{fig:Reciprocity}f shows a tight-binding lattice of trimers of cavities whose acoustic capacitance is modulated in time\cite{fleury2016floquet} exhibits sidebands as a repetition of the band structure along the frequency dimension. Floquet modulations can also result in wavenumber gaps, rather than frequency gaps for their spatial counterparts.
These host both amplified and damped modes, that can exist because of the lack of energy conservation (Sec.~\ref{breaking_energy_conservation}), as investigated in the context of spatial filtering\cite{trainiti2019time}.

It is possible to break time-reversal invariance of Floquet systems by choosing a $H(t)$ that is not an even function of time\footnote{More precisely, we want to break time-reversal invariance $\Theta H(t) \Theta^{-1} = H(-t)$ in which $\Theta$ is the time-reversal operator. When a basis can be chosen so that $H(t)$ is real and $\Theta$ acts trivially, then this constraint reduces to $H(t) = H(-t)$, but this is not always true.}. 
This is what happens in the modulated lattice of Fig.\, \ref{fig:Reciprocity}f, where the band degeneracies of the static medium are lifted upon dynamic modulation due to broken time-reversal symmetry induced by a spatially rotating phase profile\cite{fleury2016floquet}. 
This can induce non-reciprocal phonon effects with polarized coherent light\cite{disa2021engineering}.
A notable example is the phonon-mediated control of the magnetic properties of a lattice of spins, in order to induce giant paramagnetism\cite{juraschek2022giant}, phonon-driven magneto-valleytronics\cite{shin2018phonon}, or ferroelectricity\cite{radaelli2018breaking}. Slower Floquet modulations that break time-reversal symmetry can effectively impart some form of momentum and allow for strong nonreciprocal\cite{fleury2015subwavelength,shen2019nonreciprocal,chen2019nonreciprocal,wang2018observation,marconi2020experimental,chen2021efficient,shao2022electrical,wen2022unidirectional} or topological responses\cite{fleury2016floquet,darabi2020reconfigurable,chen2024observation}, or nonreciprocal acoustic devices such as robust leaky-wave antennas (Fig.\,\ref{fig:Reciprocity}f). 
{It is also possible to emulate Floquet physics through the propagation of waves in static but spatially modulated media\cite{peng2019chirality,zhu2022time,cheng2022observation}}.

\subsubsection{Time-interfaces}
\label{temporal_interfaces}
\begin{figure}[t]
\vspace*{-2cm}
\begin{minipage}[t]{0.5\columnwidth}%
\begin{center}
\includegraphics[scale=.8]{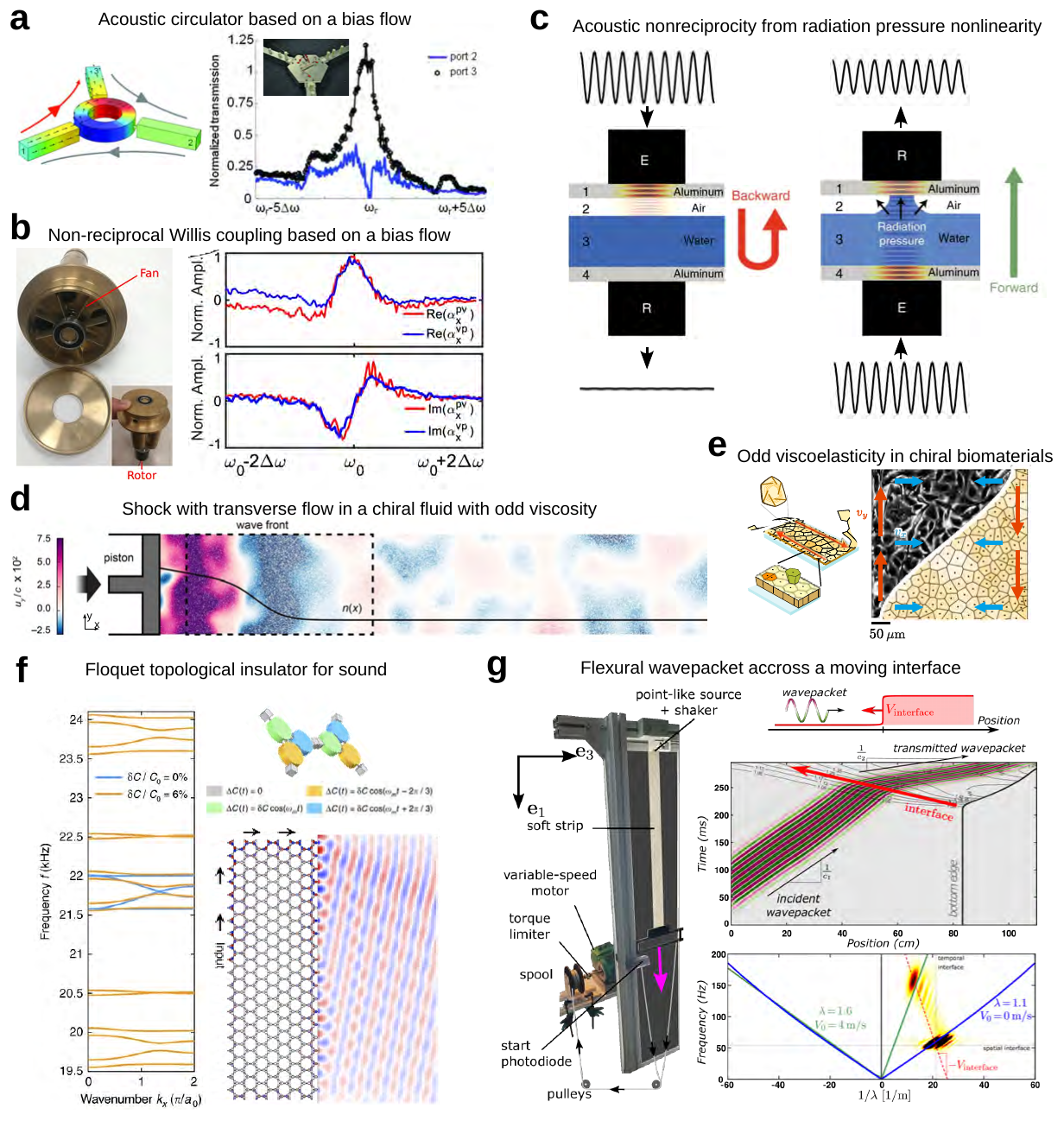}
\vspace*{-0.75cm}
\par\end{center}
\end{minipage}\hfill{}%
\begin{centering}
\centering{}
\par\end{centering}
\caption{\textbf{Phononic phenomena induced by breaking reciprocity, time-reversal and time-translation symmetry.} \textbf{a}, Acoustic circulator based on a three-port cavity (left) with embedded flow produces zero transmission from port 1 to 2, and full transmission from port 1 to 3, at the resonance frequency (right). \textbf{b}, Experimental measurement of acoustic polarizations corresponding to odd Willis coupling (right) induced by time-reversal symmetry breaking within a biased, spatially symmetrical scatterer (left). Here, the bias consists of an asymmetric flow, obtained with a motorized fan. \textbf{c}, Non-reciprocal phonon transmission induced by the combination of spatial asymmetry and nonlinearity stemming from the acoustic radiation pressure at the interface between air and water: no signal is transmitted through the device. in the backward configuration (left), while sound transmission mediated by radiation pressure in the fluid is permitted in the forward configuration (right). \textbf{d}, A shock wave propagating in a chiral fluid yields a directional transverse flow, as evidenced by the non-zero velocity $u_y$ in the direction orthogonal to the shock (colorbar). 
\textbf{e}, Chiral biological tissues exhibiting odd viscoelasticity, leading to transverse responses (red arrows) to constraints generated by cell proliferation and extrusion (blue arrows).
 \textbf{f}, Band structure of a Floquet topological insulator for sound (left) based on unit cells with a periodic time-modulation of the acoustic capacitance $C=C_0+\Delta C(t)$  at the frequency $\omega_m$ with a directional angular phase profile (top right), resulting in a topologically protected, non-reciprocal acoustic leaky-wave antenna (bottom right). \textbf{g}, Flexural wave packets crossing a spatio-temporal interface upon the abrupt deformation of a soft elastomer (left). The experiment is carried out by abruptly stretching the medium using a motor while mechanical waves are launched in the medium (top right). This results in the non-conservation of both wavenumber and frequency between the impinging and refracted wavepackets, as shown in the real space-time diagram (middle right) and frequency-wavevector diagram (bottom right), leading to different behavior depending on the excitation direction.
Panel (\textbf{a}) adapted with permission from \cite{fleury2014sound}. American Association for the Advancement of Science.
Panel (\textbf{b}) adapted with permission from \cite{quan2021odd}. Springer Nature Limited.
Panel (\textbf{c}) adapted with permission from \cite{devaux2019acoustic}. Springer Nature Limited.
Panel (\textbf{d}) adapted with permission from \cite{han2021fluctuating}. Springer Nature Limited.
Panel (\textbf{e}) adapted from \cite{chen2025chirality} with permission from the authors. 
Panel (\textbf{f}) adapted with \phantom{adapted} permission from \cite{fleury2016floquet}. Springer Nature Limited.
Panel (\textbf{g}) adapted with permission from \cite{delory2024elastic}. American Physical Society.
}
\label{fig:Reciprocity}
\end{figure}

Recent works have been interested in the temporal analogues of spatial interfaces, which induce novel scattering wave phenomena that emerge when sudden, non-adiabatic changes occur to the properties of a medium without breaking spatial-translation invariance. For example, a temporal interface can be induced in a uniform medium in which the spatially uniform refractive index is suddenly switched from one value to another. Temporal reflections and transmission emerge at such temporal interfaces, with associated temporal Fresnel coefficients and a conservation of overall momentum instead of frequency and energy, due to broken time-translation invariance but preserved spatial-translation symmetry\cite{caloz2019spacetime,liberal2024spatiotemporal}. Specifically, the incident wave is time-reversed upon temporal reflection (negative frequency), in stark contrast with the conventional mirrored spatial reflection (negative wavevector).  Various works have investigated these phenomena, both theoretically and experimentally in 1D and 2D\cite{bacot2016time,bacot2019phase,mouet2023comprehensive,apffel2022experimental}. A landmark example is the re-focusing of water waves at the surface of a basin undergoing a rapid change in gravity\cite{bacot2016time}, effectively behaving as an instantaneous version of a time-reversal mirror \cite{Fink2000}.
Soft elastomers\cite{lanoy2020dirac,delory2022soft}, whose material properties depend on the medium deformation\cite{delory2023guided}, are also a promising platform to implement experimentally spatio-temporal interfaces for phononic waves\cite{delory2024elastic}, in which a spatial interface travels at a finite velocity. The result of this interface breaking both spatial and temporal translation invariance are shown in Fig.~\ref{fig:Reciprocity}g, demonstrating the nonreciprocal conversion of both wavenumber and frequency for a wave packet impinging across the interface.

\subsection{Breaking energy conservation}
\label{breaking_energy_conservation}

On a fundamental level, energy conservation is a consequence of time-translation invariance. 
In this section, we focus on systems where energy is not conserved, but that are effectively described by a time-independent equation (contrary to Sec.~\ref{breaking_time_translation_invariance}).
In this case, energy conservation corresponds to the fact that some operators are Hermitian or anti-Hermitian.
Consider the coupled mode theory Eq.~\eqref{cmt}.
If the energy $E$ of the waves is proportional to $\lVert a \rVert^2 = \sum_m a^*_m a_m$ (where the star represents complex conjugation), then we find that $\partial_t E \propto \partial_t \langle a, (L + L^\dagger) a \rangle$ in which $\langle \cdot, \cdot \rangle$ is a Hermitian inner product and $\dagger$ represents the conjugate transpose, and so energy is conserved when $L = - L^\dagger$ i.e. when $L$ is anti-Hermitian. 
When it is the case, the eigenvalues of $L$ are purely imaginary, and correspond to the frequency of oscillation of the modes.
Equivalently, it means that the Hamiltonian $H = i L$ is Hermitian.
Conversely, systems with loss and gain are described by non-Hermitian (or non-anti-Hermitian) operators, leading to various properties common to these systems\cite{Ashida2020}.
In this section, we discuss how a careful engineering of energy balance through gain and losses can be used to tailor wave propagation in lossy and active systems.

\subsubsection{Balancing losses with gain in PT-symmetric systems}
\label{PT_symmetry}

Systems in which the eigenvalues of the Hamiltonian $H$ (the frequencies of oscillation) are purely real are called pseudo-Hermitian\cite{bender1998real,Ashida2020} or $PT$-symmetric. 
This includes lossless systems in which $H = H^\dagger$, but also systems where gain and loss are present but balanced.
The label $PT$ originally refers to the combination of parity $P$ (space inversion) and time-reversal $T$, but it turns out that a more general class of systems exhibit the same mathematical properties~\cite{Bender2002,Mostafazadeh2015}, for which there is an antiunitary operator $PT$ with $(PT)^2 = 1$ such that $PT H = H PT$.
In systems with (generalized) $PT$-symmetry, energy is not conserved in general, but when $PT$-symmetry is not spontaneously broken, it is effectively conserved when the system oscillates in a single eigenmode. 
A related symmetry known as Anti–Parity-Time ($APT$) symmetry (in which $PT H = - H PT$) has also been considered to control heat transfer~\cite{Li2019}.
When $H = i L$ (Sec.~\ref{continuum_or_else}), $H$ is $PT$-symmetric if an only if $L$ is $APT$-symmetric, and conversely\cite{Peng2016}.

For example, consider a lossy resonator with complex eigenfrequency $\omega_0 +i \gamma$ and couple it to an identical resonator with gain. We may aim at compensating the decay in the first resonator by choosing the amplification rate of the second one to be exactly equal to the loss rate of the first, i.e., with intrinsic eigenfrequency $\omega_0-i \gamma$. If we denote with $\kappa$ the rate of energy coupling between them, the Hamiltonian becomes
\begin{equation}
H =
\begin{pmatrix}
\omega_0+i \gamma & \kappa\\
\kappa & \omega_0-i \gamma\\
\end{pmatrix}.
\label{eq:coupledosc}
\end{equation}
The corresponding eigenvalues of $H$ are $\omega_0 \pm \sqrt{\kappa^2-\gamma^2}$ and the eigenvectors are proportional to $[i \gamma \pm \sqrt{\kappa^2-\gamma^2}, \kappa]^T$. In the weak coupling limit where $\kappa$ is small, the system supports two distinct modes, with complex conjugate eigenvalues, close to the ones of the individual resonators. One mode is mostly localized in the gain resonator, and it grows in time, while the other decays in time at the same rate, as it mostly resides in the lossy resonator. An interesting phenomenon happens in the opposite regime of strong coupling, where $\kappa > \gamma$. When the energy has time to circulate back and forth between the resonators before decay or gain happens, gain and loss effectively compensate each other, and the eigenvalues of the system (although described by a non-Hermitian matrix), are purely real, i.e., the modes do not grow or decay in time. This demonstrates the possibility of engineering non-Hermiticity and symmetries in a useful way, and the opportunities provided by $PT$-symmetry in the design of resonant phononic systems\cite{zhu2014p,christensen2016parity,fleury2016parity}.

The eigenvectors of non-Hermitian matrices no longer necessarily form a complete basis, and are no longer orthogonal. 
Extreme cases occur where two eigenvectors become colinear, and share the same eigenvalue. 
Such a situation is known as an exceptional point\cite{shi2016accessing,ding2016emergence,ding2018experimental,tang2020exceptional,tang2021direct,tang2023realization}.
This is indeed what happens when $\kappa = \gamma$ in Eq.~\eqref{eq:coupledosc}, for which $H$ is not diagonalizable.
The coalescence of the modes associated with the exceptional point has been evidenced in a two-level system consisting of two tightly-coupled acoustic cavities with controllable asymmetric dissipation\cite{ding2016emergence} (Fig.\, \ref{fig:EnergyCons}a).
Moreover, the form and topology of the Riemann sheets formed by the eigensurfaces around an exceptional point has sparked various studies, such as the effect of dynamic encircling of EPs\cite{tang2020exceptional,elbaz2022jphysd} to control the mode transmission, leading to optimized sound absorption\cite{achilleos2017non}, or the use of exceptional points for sensing\cite{shmuel2020prapplied,Wiersig2020}.

\subsubsection{Scattering in $PT$-symmetric systems}
\label{Compensating scattering by engineering loss and gain}

The scattering matrix $S$ (BOX2) of a lossless system is unitary, and in the steady-state the outgoing power is always equal to the incident power. 
Gain and loss affect this property and make the scattering matrix non-unitary. 
In particular, $PT$-symmetric systems (Sec.~\ref{PT_symmetry}) can restore the flux conservation when operated under proper conditions (just like $PT$-symmetric cavities show no decay in the strong coupling limit, when operated at a single eigenmode). 
In the case of scattering, $PT$-symmetric systems can support anisotropic transmission resonances\cite{ge2012conservation}, namely situations where the system is totally transparent from one side, like a lossless system, yet it can strongly reflect from the opposite side. 
This response has been experimentally realized in a two-port acoustic system with active components\cite{fleury2015invisible}, as shown in Fig.\, \ref{fig:EnergyCons}b. Under this condition, gain does not simply compensate losses ($S_{12}=S_{21}=1$), but it also cancels the reflection of the system from the lossy port ($S_{11}=0$), which is interesting from a practical standpoint. 
Excitation of the system from the opposite side yields strong reflections, despite supporting full transmission because of reciprocity, which is possible because energy conservation is broken by the presence of gain and loss.
By extending this concept, it is possible to turn a complex disordered acoustic system, initially opaque, into a completely transparent one, by adding a distribution of gain and loss tailored to counteract the arbitrary impedance fluctuations initially present\cite{rivet2018constant}. 
This directly results in a constant-pressure sound wave within the inhomogeneous sample consisting of a one-dimensional array of active acoustic scatterers (Fig.\, \ref{fig:EnergyCons}c). 
Some other potential applications related to cloaking\cite{li2019ultrathin} and directional sound emitters\cite{magariyachi2021pt} have also been explored.

\subsubsection{Virtual gain/loss through complex frequency excitations}
\label{complex_frequencies}
The presence of gain or loss in a medium implies the respective temporal growth or decay of the waves propagating in it. These aspects are associated with zeros and/or poles of the scattering matrix that are located outside of the real axis in the complex frequency plane, making them not reachable with a monochromatic excitation whose frequency is purely real. 
The usual way to circumvent this problem amounts to engineer the physical gain and loss in the system so as to move these poles and zeros towards the real frequency axis\cite{krasnok2019anomalies}.
Alternatively, one can also directly access the complex poles and zeros by exponentially increasing or decreasing in time the amplitude of waves used to probe the system.
This effectively makes the operating frequency complex\cite{kim2025complex} and allows the excitation of scattering features of the medium beyond what is possible with a purely monochromatic signal on the real frequency axis. Doing so, it is possible to make passive systems effectively behave as if they have controllable gain or loss, without the inherent issues of stability and energy consumption associated to active protocols which can make their practical realization challenging. First introduced in photonics\cite{baranov2017coherent,kim2022beyond,ra2020virtual}, these concepts of \textit{virtual} gain and loss have been applied to various non-Hermitian wave phenomena. For instance, virtual coherent absorption of elastodynamic waves has been achieved using counterpropagating signals exponentially growing in time, with a growth rate matching the leakage of the resonant inclusion \cite{trainiti2019coherent,rasmussen2023lossless}.
Besides, a temporally decaying signal has been used to implement the transient version of the non-Hermitian skin effect\cite{gu2022transient}. If applied to lossy superlenses, this virtual gain permits to effectively compensate the dissipation in the system, which enables the recovery of their full subwavelength imaging potential, uniquely limited by nonlocal constraints\cite{kim2023loss}, as shown on Fig.\,\ref{fig:EnergyCons}d.

\subsubsection{Amplification and nonlinearities}
\label{nonlinearities}

The presence of gain in a linear system eventually triggers nonlinearities that either stop the growth or destroy the system.
The poster child of nonlinear systems in wave physics is the appearance of a limit cycle, meaning that the systems starts oscillating by itself.
This can be illustrated using coupled mode theory (BOX2) by the equation
\begin{equation}
  \dot{a} = i \omega a - \gamma a + \left( \frac{\alpha^2}{\alpha + \beta |a|^2} \right) a \simeq i \omega + (\alpha - \gamma) a - \beta |a|^2 a + \mathcal{O}(a^3)
\end{equation}
for a single complex mode $a(t)$. 
The second term in this equation represents saturable gain with amplitude $\alpha$, that can be Taylor-expanded at small $|a|$ so we recognize the equation of a Stuart-Landau oscillator\cite{Kuramoto1984}: when $\alpha > \gamma$, the state $|a| = 0$ becomes linearly unstable and a limit cycle describing spontaneous oscillations of the form $a(t) = \sqrt{(\alpha - \gamma)/\beta} e^{i \omega t + \varphi_0}$ appears. 
This is the basis of operation of a laser. 
Indeed, phonon lasers, the mechanical equivalent of light lasers, have been proposed and realized in several platforms\cite{Vahala2009,Grudinin2010,Jing2014,Jiang2018,Zhang2018,Behrle2023}.
Limit cycles, which can occur because of the presence of gain and loss, spontaneously break time-reversal (in addition to time-translation invariance). 
This can be harnessed to produce nonreciprocal scattering responses with reduced losses\cite{pedergnana2024loss,Pedergnana2023}.
For instance, a sonic circulator based on spinning aeroacoustic limit cycles was recently proposed\cite{pedergnana2024loss}. As shown in Fig.\, \ref{fig:EnergyCons}e,  an air flow orthogonal to the whistle cavity provides some gain that drives the system into a limit cycle whose sustained radiation synchronizes with the incident wave to compensate for absorption losses.
More generally, nonlinear phononic media can exhibit a variety of phenomena related to underlying broken energy conservation and the possible spontaneous breaking of time-translation, ranging from nonlinear travelling waves like shocks and solitons to chaotic attractors\cite{Patil2021,Manktelow2017,Lapine2014}.

\subsubsection{Non-Hermitian skin effect}

The non-Hermitian skin effect (NHSE) is a feature of certain non-Hermitian systems~\cite{hatano1996localization,Ashida2020,Kawabata2019,Hu2024}, where eigenvalues and eigenmodes are highly sensitive to boundary conditions.
Physically, it leads to the directional amplification/attenuation of waves within the medium and to field accumulation on some specific boundaries.
In its simplest instantiation, the NHSE can be illustrated by a one-dimensional chain of asymmetrically coupled resonators similar to the quantum Hatano-Nelson model~\cite{hatano1996localization,Hu2024}. 
In phononics, this can be captured by the coupled-mode equations
\begin{equation}
  \partial_t a_n = i [ \omega a_n + t_{+} a_{n+1} + t_{-} a_{n-1} ]
\end{equation}
where integers $n$ label the resonators, and where the amplitudes of the couplings $t_\pm$ to the left and to the right are different, mimicking a biased random walk.
This effect has been related to the point gap topology of the corresponding non-Hermitian Hamiltonian\cite{ding2022non}.
In phononic media, NHSE has been investigated in the presence of asymmetric couplings \cite{ghatak2020observation,gao2020anomalous,zhang2021acoustic,zhang2021observation,wang2022non,wang2023experimental}, odd elasticity\cite{scheibner2020non,wang2024non} and odd mass densities\cite{wu2023active}. 
Among these designs, some use electronic feedback loops between microphones and speakers to obtain  NHSE for sound\cite{zhang2021observation}, shown in Fig.\,\ref{fig:EnergyCons}f which shows the accumulation of the acoustic signal at a single boundary, whatever the source position.
The NHSE can also be observed in waves reflected by lossless topological systems, where it can lead to nonreciprocal Goos-Hänchen shifts\cite{franca2022non}.

%
\subsubsection{Active continuum phononic media: odd elasticity and beyond}
\label{odd_elasticity}

\begin{figure}[t]
\begin{minipage}[t]{0.5\columnwidth}%
\begin{center}
\vspace*{-1.25cm}
\includegraphics[scale=.82]{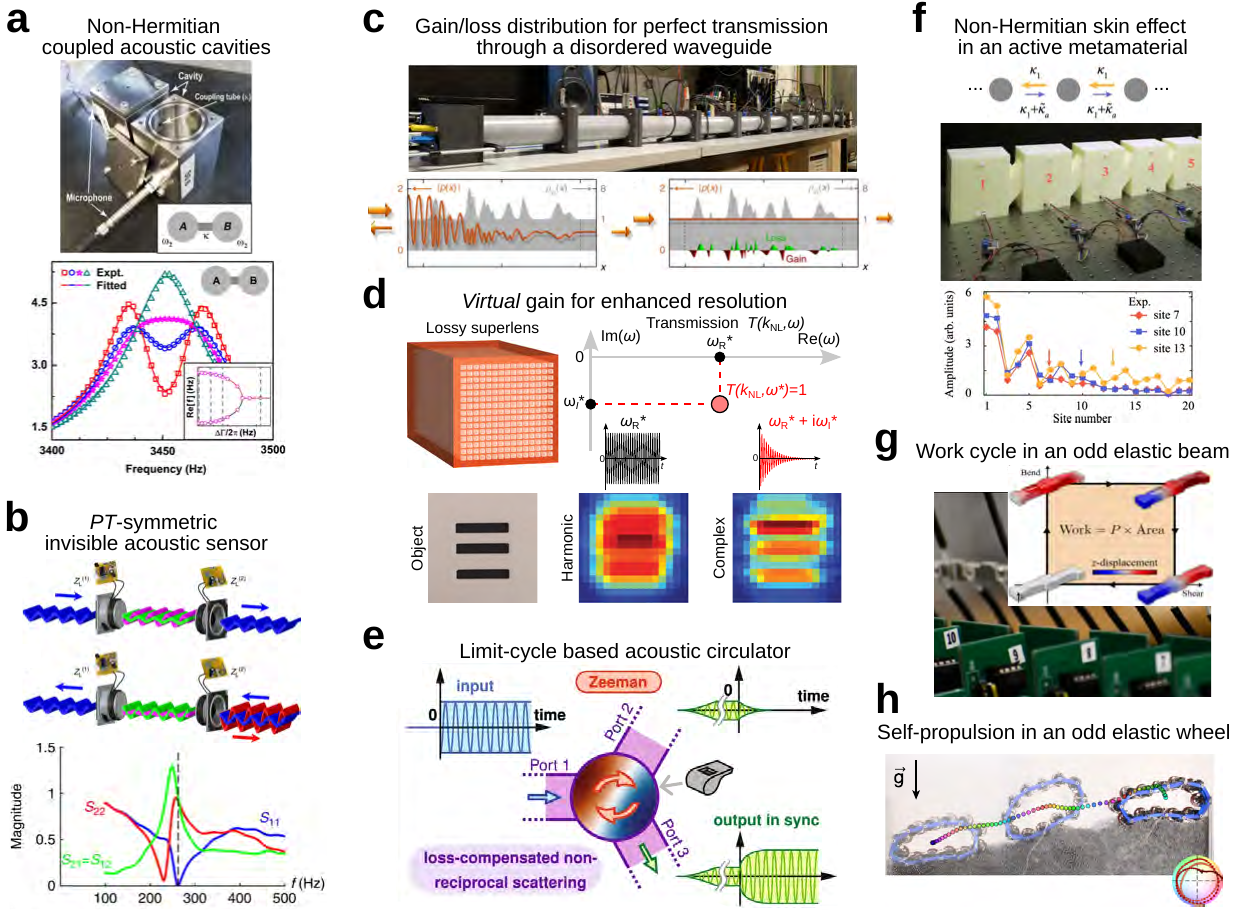}
\par\end{center}
\end{minipage}\hfill{}%
\begin{centering}
\centering{}
\par\end{centering}
\caption{\textbf{Phononic phenomena induced by breaking energy conservation.} \textbf{a}, A two-level non-Hermitian system made of two tightly-coupled acoustic cavities with controllable asymmetric loss (top) which allows for the demonstration of the coalescence of the two modes at the exceptional point, evidenced by the merging of the  two transmission peaks depending on the loss amount (bottom). \textbf{b}, An invisible acoustic sensor based on $PT$- symmetry, showcasing unitary transmission ($S_{12}=S_{21}=1$) which is reflection-less from one side ($S_{11}=0$), and with strong reflection from the other side (bottom, at gray dotted line frequency). Here, gain and loss are obtained with impedance circuit design. \textbf{c}, (Bottom left) In a disordered Hermitian acoustic system, part of the incident signal is reflected due to spatial variations of the medium properties (gray area). (Bottom right) a tailored gain-loss distribution permits to obtain a perfect transmission through the sample, along with no pressure variations within the system. (Top) this non-Hermitian design has been implemented discretely using electrodynamic loudspeakers with a controlled acoustic impedance. 
\textbf{d}, The resolution of a conventional acoustic superlens is limited by its nonlocality, $k_\mathrm{NL}$ being the associated largest wavenumber accessible, and by inherent dissipation which pushes the corresponding unitary transmission $T(k_\mathrm{NL},\omega^*)=1$ below the real frequency axis. Using the associated complex excitation ($\omega^*={\omega_\mathrm{R}}^*+i{\omega_\mathrm{I}}^*$), whose imaginary part is related to the losses of the system, instead of the monochromatic signal (${\omega_\mathrm{R}}^*$), results in a virtual gain that effectively compensate the losses in the lens and leads to enhanced resolution. 
\textbf{e}, A three-port nonlinear acoustic cavity hosts a limit cycle which continuously radiates a signal through the ports that synchronizes with the incident harmonic wave, thereby gaining energy from the limit cycle’s emissions. In the presence of a bias, this enables loss-compensated non-reciprocal transmission. \textbf{f}, Non-Hermitian skin effect in a chain of active acoustic resonators based on asymmetric couplings, which are implemented using active feedback loops between speakers and microphones.
The bottom plot shows the directional field accumulation towards the left side of the chain, whatever the source position.
 \textbf{g}, Odd properties of the elastic response of a mechanical beam induced by using electrically controlled piezoelectric patches (bottom). In such a medium with an odd micropolar modulus $P$ describing the asymmetric coupling between bending and shearing of the beam, a quasistatic cycle between bending and shear motion of the unit cell is associated with a non-zero work per unit volume whose sign depends on the direction of the cycle (top). 
 \textbf{h}, Odd elastic mechanical systems can harness their work generating cycles to produce emergent active functionalities as illustrated by this adaptive wheel that displays propulsion and uphill locomotion on a granular bed.
 Its propulsion is driven by odd couplings between two of its shear modes $S_1$ and $S_2$, whose shear-space trajectory is a noisy limit cycle (inset).
Panel (\textbf{a}) adapted with permission from \cite{ding2016emergence}. American Physical Society.
Panel (\textbf{b}) adapted with permission from \cite{fleury2015invisible}. Springer Nature Limited.
Panel (\textbf{c}) adapted with permission from \cite{rivet2018constant}. Springer Nature Limited.
Panel (\textbf{d}) adapted with permission from \cite{kim2023loss}. American Physical Society.
Panel (\textbf{e}) adapted with permission from \cite{pedergnana2024loss}. Springer Nature Limited.
Panel (\textbf{f}) adapted with permission from \cite{zhang2021acoustic}. Springer Nature Limited.
Panel (\textbf{g}) adapted with permission from \cite{chen2021realization}. Springer Nature Limited.
Panel (\textbf{h}) adapted with permission from \cite{Veenstra2025}. Springer Nature Limited.}

\label{fig:EnergyCons}
\end{figure}

We have seen in Sec.~\ref{non_reciprocal_continuum} that the elastic tensor in Eq.~\eqref{eq:EffConstRel} violates reciprocity when $C_{ijk\ell}\neq C_{k\ell ij}$.
Similarly, energy conservation is violated whenever $C_{ijk\ell} \neq C_{k\ell ij}^*$ ($*$ represents complex conjugation\cite{Muhlestein20160Prsa2}, both constraints match when $\elas$ is real-valued).
Such a situation, known as odd elasticity, can arise from the presence of non-conservative microscopic interactions\cite{scheibner2020odd,fruchart2023odd} and has been realized with active elements\cite{chen2021realization,Veenstra2025}, as in the examples of Fig.\,\ref{fig:EnergyCons}g showing a beam decorated with piezoelectric patches with designed feedback loops\cite{chen2021realization}.
In such a metamaterial, quasistatic cycles between the bending and shear modes are associated with a non-zero amount of work whose sign depends on the cycle directionality. Such cycles could have implications for energy harvesting and sensing, and enable active wave propagation and instabilities within overdamped media \cite{scheibner2020odd,scheibner2020non,fossati2024odd} as well as adaptive locomotion of active solid metamaterials\cite{Veenstra2025} (see Fig.\,\ref{fig:EnergyCons}h).
Conversely, the presence of these cycles mean that it is impossible to have odd elasticity in a system where energy is conserved.
Similarly, constraints on other material coefficients (like $\rho_{ij}$ and $S_{ijk}$) due to energy conservation~\cite{Muhlestein20160Prsa2} can be broken in media where energy is not conserved\cite{wu2023active,quan2021odd}.

Besides, when $C_{ijk\ell}$ depends on frequency, it can also include a viscous part that dissipates energy, but can be passive.
The combination of both leads to a viscoelastic medium\cite{Lakes1998,lakes2009viscoelastic} in which $C_{ijk\ell}(\omega) = C_{ijk\ell}(0) + i \omega \, \eta_{ijk\ell}(0) + \mathcal{O}(\omega^2)$ where $C_{ijk\ell}(0)$ and $\eta_{ijk\ell}(0)$ are the the (zero-frequency) elastic and viscous tensors.


\section{Generalized symmetries}
\label{generalized_symmetries}

In this section, we discuss several situations in which it is fruitful to consider the symmetries of multiple systems at once.
In section \ref{symmetry_to_topology}, we discuss how the interface between two systems that have different symmetries (or different representations of the same symmetry) can host topological surface states. 
In section \ref{duality}, we discuss how generalizing symmetries to families of systems depending of parameters can allow us to identify additional hidden symmetries and to construct isospectral crystals.
In section \ref{nonlocality_engineering}, we discuss how interpolating between different breakings of spatial-translation invariance can be used to enhance the control of the dispersion relation of phononic crystals.
In section \ref{twistronics}, we discuss how multi-layers systems in which the symmetries of each layer can be harnessed to tailor wave propagation.

\subsection{From symmetry to topology}
\label{symmetry_to_topology}

Topological band theory is a framework that harnesses tools from topology to understand and control the behavior of waves in materials, including their interfacial response (see Sec.~\ref{Interfaces} and references therein).
Symmetry plays a key role in the design of topological phononic crystals.
In a nutshell, edge states often occur when the symmetry of states on both sides of an interface do not match. 
More precisely, consider a system depending on a parameter $p$ so that a band crossing occurs where two irreducible symmetric representations cross each other at some critical value $p_c$. At an interface between $p<p_c$ and $p>p_c$, a band crossing is susceptible to occur to interpolate between the band structures on both sides.
In such a system, there is a band gap on both sides of the interface (when $p \neq p_c$), but the gap closes at the interface.
The existence of such an edge state is to some extent unavoidable and related to the topology of the band structure.
This example illustrates a general principle that has been formalized in group-theoretical terms under the name of topological quantum chemistry\cite{bradlyn2017topological,cano2021band}, allowing to obtain a complete catalog of topological phononic media\cite{xu2024catalog}.

The lifting of band degeneracies through symmetry-breaking is a common starting point for creating band-gaps and topological boundary modes. 
This is for instance the case of Dirac cones in honeycomb lattices, singularities protected by time-reversal and inversion symmetries. 
Breaking time-reversal yields a Chern topological insulator, exhibiting non-reciprocal phononic propagation at the edge that is robust against spatial disorder\cite{khanikaev2015topologically,ding2019experimental} (Fig.\, \ref{fig:GlobalSymm}a).
Breaking spatial inversion symmetry while preserving time-reversal also opens a band-gap, leading to a valley-Hall insulator\cite{lu2017observation}. 
There, the edge modes between two mirrored lattices are only robust when a pseudo-spin associated to the two valleys is conserved.
This emergent pseudo-spin can be harnessed to endow phononic waves with effective fermionic properties.
This idea has been extensively investigated in phononics across multiple platforms such as lattices of pillars with different radii\cite{he2016acoustic}, which allows for pseudo-spins based on $C_{6v}$ symmetry and dipolar and quadrupolar modes, whose conservation allows for wave sorting (Fig.\, \ref{fig:GlobalSymm}b). 
A similar symmetry-based strategy has been proposed to self-assemble topological phononic metamaterials~\cite{Fruchart2018}.

\subsection{Duality for phonons}
\label{duality}
The general definition of a symmetry, a transformation that leaves a system invariant, leaves room for other kinds of symmetries beyond the ones discussed in Secs.~\ref{breaking_spatial_symmetries} and \ref{breaking_nonspatial_symmetries}, that can be less straightforward, or somewhat hidden. 
An example is given in Fig.\,\ref{fig:GlobalSymm}c, where two different mass-spring chains share an identical band structure at the scale of the Brillouin zone, albeit the absence of any conventional symmetrical relation between them. These “hidden symmetries” may look at first glance to be accidental. To understand their nature, it is convenient to consider an entire family of systems continuously depending on a parameter $p$ rather than looking at each system individually. Formally, to define a symmetry, the structure of the system of interest needs to be encoded into a mathematical object such as the operator $L$ in Sec.~\ref{continuum_or_else}, a Hamiltonian, or a dynamical matrix.
Transformations are encoded into operators $U$, that are symmetries of $L$ provided that $U L U^{-1} = L$.
In this case, it is implicit that parameters are the same in $L = L(p)$ on both sides of the equation.  
Dualities can be seen as symmetries that also change the parameter $p$ associated to the system: given a function $p \mapsto u(p)$ acting on parameters, we say that $U$ is a duality when $U L(p) U^{-1}=L(u(p))$.
At fixed points $p^*$ of $u$ such that $u(p^*)=p^*$, known as self-dual points, the duality reduces into a conventional symmetry operation. 
The Onsager-Casimir reciprocal relations\cite{nassar2020nonreciprocity} relating the properties of systems with opposite external biases (e.g. magnetic fields, see Sec.~\ref{breaking_reciprocity}) are an example of such a duality.
The mechanical structure known as a twisted Kagome lattice\cite{fruchart2020dualities} shown in Fig.\,\ref{fig:GlobalSymm}d provides an example with a more complex duality operator.
In this case, the parameter $p=\theta$ is the twist angle, and $H$ stands for the dynamical matrix of the phonons. 
The duality operator $U$ shuffles the vibrational degrees-of-freedom in the unit cell as shown in Fig.\,\ref{fig:GlobalSymm}d (top inset), and $u(p)=-p$. Because of the duality within the family of twisted lattices, the band structures of dual systems are identical. 

Other examples of mechanical systems dualities can be found in\cite{fruchart2023systematic,lei2022duality,yang2023non}, including a systematic way to construct them\cite{fruchart2023systematic}. In addition, the abstract Maxwell duality between floppy modes and states of self-stress\cite{zhou2019topological,kane2014topological,mcinerney2020hidden,czajkowski2024duality}, can also translate into a physical duality between parallelogram tilings and fiber networks\cite{zhou2019topological}. Beyond the iso-spectrality of dual media, duality operations put additional constraints on the self-dual configuration, similarly to conventional symmetries. More generally, wave dualities have consequences on the macroscopic elastic properties\cite{fruchart2020symmetries}, affect the propagation of waves at interfaces\cite{gonella2020symmetry} as well as  topological edge and corner states\cite{danawe2021existence,azizi2023dynamics,allein2023strain}. They also lead to pseudo-spin degeneracies unusual in mechanics that can be exploited to perform information processing using non-Abelian geometric phases\cite{fruchart2020dualities}.

\subsection{Using families of symmetries to control phononic band structures} 
\label{nonlocality_engineering}

The crystallographic approach based on spatial symmetries is a powerful tool to engineer phononic crystals, but as the number of possible combinations of symmetries (space groups) is limited, this approach is not sufficient to finely control the shape of the bands.
One way to circumvent these limitations is to use a parameter-dependent material, like in Sec.~\ref{duality}, in order to interpolate between different combinations of symmetries\cite{chen2021roton,bossart2023extreme}.
In particular, nonlocal metamaterials use couplings going beyond nearest-neighbors to control the propagation of waves\cite{kazemi2023drawing,moore2023acoustic,chaplain2023reconfigurable,chen2021roton}.
Consider the simple case of the monoperiodic chain of meta-atom with families of couplings corresponding to different spatial ranges\cite{chen2021roton}, as shown in Fig.\,\ref{fig:GlobalSymm}e. 
There, the system with blue couplings only or red couplings only correspond to different discrete translation symmetries.
While the periodicity of the system remains constant, the band curvature strongly depends on the ratio of the couplings strengths. This behavior can also be seen as the result of the interaction of several chains of meta-atoms with different periodicities, which allows for multiple mode scales at a fixed frequency in the first Brillouin zone. In particular, the selective promotion of the third order inter-cell couplings generates a dispersion relation with a local minimum that mimics the physics of rotons in superfluids.
This roton-like dispersion has been experimentally implemented in 3D phononic metamaterials\cite{chen2021roton,iglesias2021roton,wang2022nonlocal,zhu2022observation,chen2023phonon,chen2023observation} and allows the propagation of multiple traveling waves with different wavenumbers and opposite directions at the same frequency in a homogeneous material, as well as zero-group velocity modes at the two inflection points.
A similar strategy has been used to design delocalized zero-energy modes with a frequency $\omega = 0$ at a tunable wavevector $\bm{q} \neq \bm{0}$ by inducing nontrivial rigid motions within the medium using graph theoretical tools\cite{bossart2023extreme}.
The hybridization of these modes with the waves propagating in the lattice generates anomalous cones in the dispersion relation emerging from arbitrary locations in the Brillouin zone, different from what conventional band folding can achieve (see Fig.\,\ref{fig:GlobalSymm}f), and leading to broadband negative refraction.


\subsection{Twistronics for phonons}
\label{twistronics}

\begin{figure}[t]
\vspace*{-1.25cm}
\begin{minipage}[t]{.5\columnwidth}%
\begin{center}
\includegraphics[scale=.83]{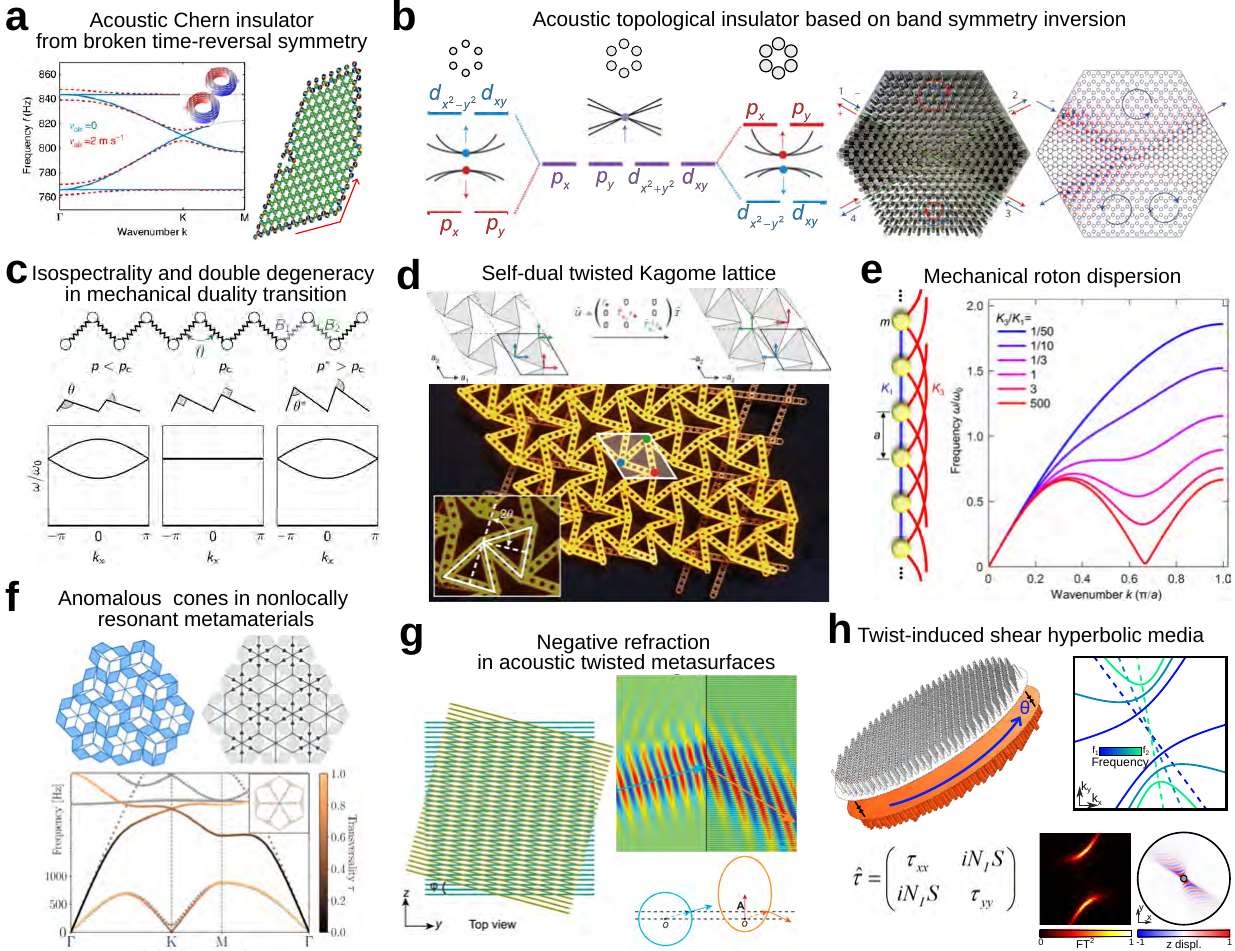}
\par\end{center}
\end{minipage}\hfill{}%
\begin{centering}
\centering{}
\par\end{centering}
\caption{\textbf{Generalized symmetries within families of phononic media.}
 \textbf{a}, Flow-induced time-reversal symmetry breaking in an hexagonal acoustic lattice lifts the Dirac cone degeneracy and open a topological bandgap described by a Chern invariant. Such a phononic Chern insulator exhibits topologically protected non-reciprocal wave propagation at is boundaries. \textbf{b}, The preservation of $C_{6v}$ spatial-symmetry within triangular lattices of pillars, as well as the symmetrical band inversion induced by changing the radius of the pillars (left), results in helicoidal topological boundary states whose symmetry-engineered pseudo-spin allows for wave sorting at topological crossings (right). \textbf{c}, Three different configurations of a mass-spring chain show the emergence of a hidden symmetry, or duality, that results in two different lattices with the same band structure (left and right). At the self-dual point (center), the duality becomes a symmetry of the medium, and leads to a double degeneracy of the band for the entire Brillouin zone. \textbf{d}, In the plane, a twisted Kagome lattice hosts a duality transformation $U$ that rotates the displacements of that relates the orientation of the three masses (red, blue and yellow arrows in the schematic at the top) and translates them to different unit cells.
 This non-trivial duality relates different twisted Kagome lattices configurations (top left and right lattices), which can be implemented using LEGO bricks (bottom).
\textbf{e}, Bending of the dispersion relation in an unidimensional chain of masses and springs as a function of the ratio between first-order and third order couplings. For a ratio equal to 1, the dispersion is analogous to that of a roton. 
\textbf{f}, Zero-energy deformation mode within a nonlocally resonant phononic metamaterial (top left) with its graph description on the underformed lattice (top right), which yields anomalous cones in the band structure emerging from zero frequency at the corner K of the Brillouin zone (bottom), in contrast with the typical case of cones starting at the $\Gamma$ point. 
\textbf{g}, The twist between phononic crystal layers (left) induces a gauge field leading to negative refraction of acoustic waves (top right), described by a shift of the isofrequency contours in the reciprocal space, which can be modeled by a spatially-dependent gauge field $A$ (bottom right). \textbf{h}, The twist between two detuned anisotropic elastic metasurfaces (top left) yields a hyperbolic effective material tensor $\bm{\tau}$ with non-Hermitian off-diagonal terms (bottom left). This leads to a frequency dependency of the hyperbolic contour orientation of the bilayer (top right). This axial dispersion comes with an asymmetric distribution of losses, which results in shear hyperbolic wavefronts of the out-of-plane displacement stemming from a point source excitation, as shown in both reciprocal and real space (bottom right).
Panel (\textbf{a}) adapted with permission from \cite{khanikaev2015topologically}. Springer Nature Limited.
Panel (\textbf{b}) adapted with permission from \cite{he2016acoustic}. Springer Nature Limited.
Panel (\textbf{c}) adapted with permission from \cite{fruchart2023systematic}. American Physical Society.
Panel (\textbf{d}) adapted with permission from \cite{fruchart2020dualities}. Springer Nature Limited.
Panel (\textbf{e}) adapted with permission from \cite{chen2021roton}. Springer Nature Limited.
Panel (\textbf{f}) adapted with permission from \cite{bossart2023extreme}. American Physical Society.
Panel (\textbf{g}) adapted with permission from \cite{yang2021demonstration}. American Association for the Advancement of Science.
Panel (\textbf{h}) adapted with permission from \cite{yves2024twist}. American Physical Society.
}
\label{fig:GlobalSymm}
\end{figure}

In the case of multi-layer systems, it is possible to combine the spatial symmetries associated of each layer, going beyond the perspective discussed in Sec.~\ref{breaking_spatial_symmetries}, through
inter-layer rotations for instance.
Recent works have started transposing the ideas of twistronics\cite{carr2017twistronics} from electronic systems to phononics as an additional knob for wave control\cite{oudich2024engineered}. 
For example, the interplay between the spatial symmetries of the layers can generate superlattices controlled by the twist operation.
These emerging moir\'e patterns with long-range periodicity at specific twist angles are responsible for flat bands in the dispersion relation of the twisted multilayer, which are tightly linked to field localization and strong resonant behavior\cite{lopez2020flat,deng2020magic,gardezi2021simulating,marti2021dipolar,lopez2022theory}. 
Moiré patterns can also emerge within a monolayer platform made of resonators whose positions are fixed but whose resonant properties are spatially modulated and rotated to introduce a structural mismatch with the underlying resonant lattice. Such twisted spatial modulations of a unique layer yield tunable wave behavior, as demonstrated in hyperbolic phononic metasurfaces\cite{yves2022moire}.
Beyond strictly periodic moir\'e patterns, twist-driven topological effects\cite{rosa2021topological,wu2022higher} and tunable gauge fields for negative refraction\cite{yang2021demonstration} relying on structural features have been demonstrated over large angular ranges, as showcased on Fig.\,\ref{fig:GlobalSymm}g.

Beyond twist-induced lattice effect, inter-layer rotations are relevant for anisotropic media (Sec.~\ref{breaking_isotropy}) as showcased by twisted hyperbolic metasurfaces, where monolayers with hyperbolic dispersion are coupled and rotated with respect to each other\cite{hu2020moire,yves2022topological,han2024nonlocal}.
By controlling the twist angle, the bilayer system undergoes a topological transition between open and closed frequency contours for a broad range of frequencies, enabling a broadband tunability of the directionality and localization of the wave propagation. 
In particular, the transition \textit{magic-angle} corresponds to a canalization regime with enhanced wave matter interaction.
Twist effects can also be used to rotate or shear dispersion relations in so-called twisted shear hyperbolic metasurfaces\cite{yves2024twist}.
The corresponding orthogonality breaking between two detuned directional resonances makes possible to control both the Hermitian and non-Hermitian features of the wave propagation.
For a fixed twist angle, the principal axis of the hyperbolic medium rotates with frequency, and the spatial distribution of loss does not match the contour’s symmetry. 
At the operating frequency, this translates into an effective material tensor $\bm{\tau}$, whose Hermitian part is diagonal while its non-Hermitian part presents some off-diagonal terms (Fig.\,\ref{fig:GlobalSymm}h). 
Using the twist between two detuned hyperbolic metasurfaces, this effect can be maximized, and directly results in a screwed hyperbolic field profile where some branches are overdamped and others enhanced in comparison to a conventional hyperbolic medium. Besides, the combination of additional rotation symmetry breakings within multilayer phononic media yields even more advanced wave manipulation\cite{han2025observation}, such as all-angle directional canalization of sound\cite{han2025all}.  


\section*{Outlook}


As showcased throughout this review, a symmetry-driven approach turns out to be a successful paradigm for advanced manipulation of phononic fields, across a wide range of domains and scales. 
In Sec.~\ref{breaking_energy_conservation}, we have reviewed the behavior of systems where energy conservation is broken. 
In that regard, a frontier consists in engineering active and time-dependent media with feedback, in which wave propagation can be controlled at will through feedback loops.
The future development of multi-physics concepts like electro- or magneto-momentum couplings has a great potential to push the levels of reconfigurability of phononic media beyond what is currently possible. 
In particular, the ability to implement extremely fast modulations of the global properties of the medium over a large scale would allow the investigation of out-of-equilibrium physical phenomena\cite{bachelard2017emergence} and the use of time as an extra tuning parameter in the context of 4D metamaterials\cite{engheta2023four}. Reconfigurable acoustic metasurfaces paired with optimization protocols have also proven to be a crucial tool for wavefield shaping, hereby allowing advanced multiplexing of acoustic communication in complex environments\cite{bourdeloux2024solution}. Alternatively, the use of active matter\cite{Marchetti2013,shankar2022topological}, as well as flexible soft elastomers\cite{lanoy2020dirac,delory2022soft,delory2023guided} to modify the properties of the propagating medium opens the door to complex dynamic wave phenomena for which analogies with the behavior of active solids ranging from biological tissues to soft robotic materials could be drawn\cite{Marchetti2013,shankar2022topological,fruchart2023odd,Baconnier2025,Henkes2011,Baconnier2022,Gu2025,Veenstra2025}.
Focusing on the zero-frequency response of the medium\cite{jiao2023mechanical,bertoldi2017flexible}, active components opens avenues for autonomous metamaterial-based machines which are of relevance for sensing, shape-morphing and object manipulation.
Conversely, the use of activity often requires to take into account instabilities and nonlinearities (Sec.~\ref{nonlinearities}), which can be used to create new functionalities.
In that context, new ideas and implementations linked to non-Abelian, non-Hermitian and non-linear topological phenomena, as well as topological defects and disordered topological phases have emerged as a means to control acoustic and optical fields\cite{zhang2023second,yang2024non,Szameit2024}. 
These directions hold promise for next-generation computation and telecommunications applications built on topologically robust devices.

The symmetry-driven approach described in this review goes beyond artificial media, and also applies to natural materials.
In the near-infrared optical frequencies, the vibrations of atomic lattices (phonons) can interact with light to create quasi-particles called phonon-polaritons, whose symmetry-related properties, such as hyperbolicity, are currently under extensive theoretical and experimental study\cite{galiffi2024extreme}. 
Controlling the propagation of these hybrid surface waves, either via twisted multi-layer systems or artificial patterning, is at the heart of modern nano-photonics and future investigations combining both spatial and time symmetries, such as Floquet polaritonics, are promising research directions.  
Beyond phonon-polaritons, it has been recently demonstrated via inelastic X-ray scattering that phonons in natural alpha-quartz exhibit an intrinsic chirality\cite{ueda2023chiral}. 
Phonons are also related to heat transport\cite{Li2012} and the symmetry-based approach also applies to this diffusive regime, as showcased by thermal systems with anti-parity-time symmetry\cite{Li2019} and twisted thermal metasurfaces~\cite{Li2024}.

Going beyond standard symmetries, the examples of the generalized symmetries discussed in Sec.~\ref{generalized_symmetries}, such as twist symmetries (Sec.~\ref{twistronics}) or dualities (Sec.~\ref{duality}), or the example of anti-parity-time symmetry show that the symmetry-based approach discussed in this review is open-ended as new kinds of symmetries can emerge.
{Finally, the efficient design methods using inverse problems and machine learning techniques to implement artificial meta-structures
which have been developed in past decade \cite{ronellenfitsch2019inverse,mao2020designing,van2022machine,bastek2022inverting,oudich2023tailoring,maurizi2025designing} can be complemented by symmetry-based approaches, for instance by using equivariant machine learning techniques~\cite{Villar2021}}.

\noindent\textbf{Acknowledgements}\\
A.A. and S.Y. were supported by the National Science Foundation Science and Technology Center 'New Frontiers of Sound', the Department of Defense and the Simons Foundation. G.S. acknowledges funding by the
European Union (ERC, EXCEPTIONAL, Project No. 101045494). M.R.H. acknowledges support from Office of Naval Research under Award No.~N00014-23-1-2660. R.F. acknowledges funding by the Swiss National Science Foundation under the Eccellenza Award 181232.
M.F. and V.V. acknowledge partial support from the France Chicago center through a FACCTS grant. 
V.V. acknowledges partial support from the Army Research Office under grant W911NF-22-2-0109 and W911NF-23-1-0212, the National Science Foundation through the Center for Living Systems (grant no. 2317138), the National Institute for Theory and Mathematics in Biology, the Chan Zuckerberg Foundation and the Simons Foundation.

\noindent\textbf{Author contributions}\\
All authors contributed substantially to the discussion of the content. A.A.~and M.R.H.~initiated the project. S.Y., M.F., R.F., G.S., V.V.~researched the data and wrote the respective sections of the article. All authors reviewed and edited the manuscript.

\noindent\textbf{Competing interests}\\
The authors declare no competing interests.


\bibliography{BiblioMerged}


\fbox{\begin{minipage}{48em}
\noindent 
\includegraphics[width=1\linewidth]{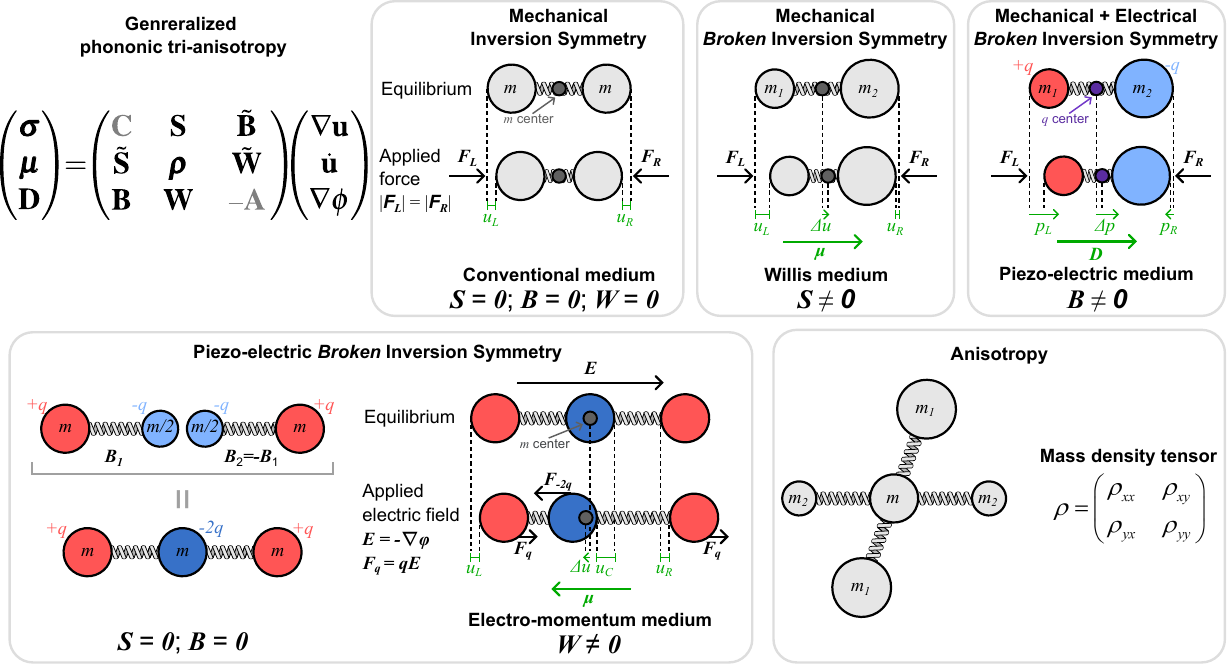}
\\
\textbf{Box 1 | Breaking spatial symmetries from microscopic to macroscopic scales.} The elastodynamic and electrostatic response of a system is described by its constitutive relations that relate kinetic (stress $\Stress$, linear momentum $\momentum$, electric displacement $\boldsymbol{D}$) and kinematic fields (displacement gradient $\boldsymbol{\grad\disp}$, velocity $\boldsymbol{\dot{u}}$, and electric field $\boldsymbol{E}=-\boldsymbol{\grad\potential}$). It is possible to generate couplings between these equations by carefully breaking the relevant spatial symmetries. In such a case, the system becomes tri-anisotropic, and can be described by a constitutive tensor whose off-diagonal constituents corresponds to coupling quantities induced by symmetry-breaking\cite{pernassalomn2020prapplied}. The 1D mass-spring model displayed in this box gives a simplified picture of the relation between spatial symmetry breaking and the associated off-diagonal terms, and a more rigorous and complete description can be found in\cite{Muhafra2023PRApplied}. Starting from the case of two identical masses linked by a spring, the application of symmetrical forces of equal amplitude on both side of the system does not change the position of the center of mass. If the masses are different, however, the center of mass moves by $\boldsymbol{\Delta u}$, which differs from the displacement of the centroid. This results in the emergence of a non-zero linear momentum upon the application of a time-varying symmetrical stress $\Stress$, described by Willis coupling $\boldsymbol{S}\neq\boldsymbol{0}$. In addition, if the different masses have opposite electric charges, the time-varying symmetrical stress results in a non-zero global electric polarization $\boldsymbol{\Delta p}$. In turn, this yields an electric displacement field $\boldsymbol{D}$, which is described by a piezoelectric coupling $\boldsymbol{B}\neq\boldsymbol{0}$. Building on this model of piezoelectric coupling, we can couple two identical dimers, asymmetric in charge and mass, in a mirrored configuration. This corresponds to spatial inversion symmetry breaking of the piezoelectric coupling itself, which can be modeled by a three mass-spring system with both mechanical and electrical spatially symmetric features, preventing the existence of global Willis $\boldsymbol{S}=\boldsymbol{0}$ and piezoelectric $\boldsymbol{B}=\boldsymbol{0}$
 couplings. Nevertheless, the presence of an electric field $\boldsymbol{E}$ results in an asymmetric motion of the three charged masses driven by the Coulomb interaction $\boldsymbol{F_q}$, which changes the position of the center of mass of the system. This leads to the emergence of electromomentum coupling, $\boldsymbol{W}$, between the applied time-varying electric potential and the linear momentum. Finally, a global anisotropy of the medium, namely different responses as a function of the excitation direction, can be embedded into a dynamical mass density tensor. While additional phenomena should be taken into account for the theoretical description of these couplings in a real material\cite{pernassalomn2020prapplied}, this simplified mass-spring model highlights the origin of the dynamic effective field-coupling properties as spatial symmetry-breaking at the microscopic scale. The sketches in this box are inspired by the work of\cite{Muhafra2023PRApplied}. American Physical Society.
\end{minipage}}


\fbox{\begin{minipage}{48em}
\noindent 
\textbf{Box 2 | Non-spatial symmetries from a scattering perspective.}
The scattering matrix $S$ summarizes how waves are reflected and transmitted (scattered) off an object\cite{Laude2009,Dwivedi2016,Schomerus2017,Markos2008}.
It has the same content as the transfer matrix (Sec.~\ref{phononic_crystals}), organized in a different fashion.
Mathematically, it relates the amplitudes $s^{\text{in}}_\beta$ and $s^{\text{out}}_\alpha$ or incoming and outgoing waves through different \enquote{channels} (labelled by $\alpha, \beta, \dots$) through $s^{\text{out}}_\alpha = S_{\alpha\beta} s^{\text{in}}_\beta$. 
The channels can be physical (e.g. waveguides) or abstract (different angles or polarizations).

In terms of the scattering matrix, the fundamental non-spatial symmetries that can be present in a phononic medium are\cite{Carminati2000,Maznev2013,Guo2022} (i) reciprocity $S = S^T$, (ii) energy conservation $S S^\dagger = 1$ and (iii) time-reversal invariance $S S^* = 1$.
These symmetries are different from each other, so it is possible to have a medium response such as reciprocity without energy conservation (e.g. in a passive lossy medium). However, any two of the symmetries implies the third, so a lossless medium that is also invariant under time-reversal must be reciprocal.

The scattering matrix associated to a set of resonant modes can be computed starting from coupled mode theory (Sec.~\ref{continuum_or_else}).
When the system of interest is coupled with the outside environment (through the channels $\alpha$), Eq.~\eqref{cmt} becomes\cite{Fan2003,WonjooSuh2004}
\begin{equation}
  \dot{a}_m = L_{mn} a_n - \Gamma_{mn} a_n + {W}_{m \alpha} s^{\text{in}}_\alpha
  \qquad
  \text{and}
  \qquad
  s^{\text{out}}_\alpha = S^{0}_{\alpha \beta} s^{\text{in}}_{\beta} + \tilde{W}_{\alpha n} a_{n}
\end{equation}
in which the operator $L$ describes the behaviour of the uncoupled system ($H = i L$ would be the Hamiltonian), $\Gamma$ and ${W}$ represent the loss and gain in the system due to exchanges of waves with the channels, respectively; $\tilde{W}$ represents the emission of waves in the channels from the resonant modes, and $S^{0}$ is the scattering matrix that the system would have if there were no resonant modes, i.e.~ when $a=0$.
Probing the system with monochromatic waves at frequency $\omega$ eventually imposes $a(t) = A e^{-i \omega t}$. 
Eliminating the resonant modes $a(t)$, we then find that the effective scattering matrix such that $s^{\text{out}} = S^{\text{eff}} s^{\text{in}}$ reads
\begin{equation}
  S^{\text{eff}} = S^0 - \tilde{W} \left[ L - \Gamma + i \omega \right]^{-1} {W}
  \label{Seff}
\end{equation}
This equation is known as the Mahaux-Weidenmüller formula\cite{mahaux1969shell}.
In a system where energy is conserved, $S^{0}$ is unitary ($S^{0\dagger} S^0 = 1$), $L$ is anti-Hermitian ($L = - L^\dagger$), and one must have $\partial_t \lVert a \rVert^2 = \lVert s^{\text{in}} \rVert^2 - \lVert s^{\text{out}} \rVert^2$, leading to a self-energy $\Gamma = 1/2 \tilde{W}^\dagger \tilde{W}$ and to ${W} = - \tilde{W}^\dagger S^0$. 
When in addition $S^{0} = 1$, the Mahaux-Weidenmüller formula then reduces to the more familiar form $S^{\text{eff}} = 1 + {W}^\dagger \left[ L - \tfrac12 W W^\dagger + i \omega \right]^{-1} W$.
In this framework, losses can be modeled by additional channels $\alpha$ with $s^{\text{in}}_\alpha = 0$ (on average), and for which one does not monitor the output $s^{\text{out}}_\alpha$. Indeed, any purely dissipative (positive-semidefinite) loss term $\Gamma$ can be modeled this way~\cite{Zirnstein2021} by adding enough ports and setting $W \propto \Gamma^{1/2}$.
From this scattering perspective, we reach three conclusions:
(i) engineering the coupling of a closed system to a radiation continuum is a way to induce effective non-Hermitian properties; (ii) the symmetries of $L_0$ and of the couplings directly influence the ones of $S$; and (iii) one can engineer
poles and zeros of $S$, and even operate near them by exciting the system with complex frequencies (Sec.~\ref{complex_frequencies}).
When the operator $L$ depends explicitly on time, then it is not possible to focus on a single frequency $\omega$ and the above formalism has to be adjusted to account for the energy transfer to frequency harmonics.
(Please note that the scattering matrix $S$ and the coupling operators $W$ and $\tilde{W}$ introduced in this BOX are different from and unrelated to the Willis and electromomentum coupling tensors ${\willis}$, $\rg$, and $\tilde{\rg}$ in Eq.~\eqref{eq:EffConstRelEM}.)
\\[1em]
\begin{center}
\includegraphics[scale=1]{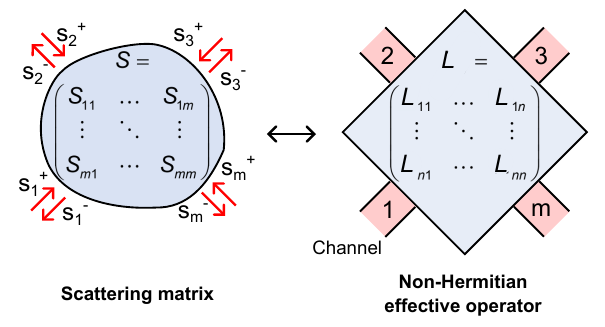}    
\end{center}

\end{minipage}}

\end{document}